\documentclass[11pt, a4paper]{article}

\usepackage[utf8]{inputenc}
\usepackage[T1]{fontenc}
\usepackage{lmodern}

\usepackage[margin=1in]{geometry}
\usepackage{setspace}
\usepackage{titlesec}
\titleformat*{\section}{\large\bfseries\boldmath}
\titleformat*{\subsection}{\normalsize\bfseries\boldmath}
\titleformat*{\paragraph}{\normalsize\bfseries\boldmath}
\usepackage{amsmath, amssymb, amsthm, amsfonts, mathtools, xspace, dsfont}
\usepackage{booktabs,microtype,needspace,placeins}
\allowdisplaybreaks

\usepackage[most]{tcolorbox}
\usepackage[colorinlistoftodos, textsize=footnotesize]{todonotes}
\usepackage[authoryear,round,longnamesfirst]{natbib}

\definecolor{darkblue}{rgb}{0.0, 0.0, 0.55}
\usepackage[pagebackref=true, colorlinks=true, allcolors=darkblue]{hyperref}
\renewcommand*{\backref}[1]{}
\renewcommand*{\backrefalt}[4]{%
    \ifcase #1%
    \or #2%
    \else #2%
    \fi%
}

\makeatletter
\newtheoremstyle{smallcaps}
  {}
  {}
  {\itshape}
  {}
  {\scshape}
  {.}
  { }
  {}

\renewenvironment{proof}[1][\proofname]{\par
  \pushQED{\qed}%
  \normalfont \topsep6\p@\@plus6\p@\relax
  \trivlist
  \item[\hskip\labelsep
        \scshape
    #1\@addpunct{.}]\ignorespaces
}{%
  \popQED\endtrivlist\@endpefalse
}
\makeatother
\theoremstyle{smallcaps}
\newtheorem{theorem}{Theorem}[section]
\newtheorem{lemma}[theorem]{Lemma}

\newcommand{\prpar}[1]{\operatorname{Pr}[#1]}
\newcommand{\prsub}[2]{\operatorname{Pr}_{#1}\left[#2\right]}
\newcommand{\prsubpar}[2]{\operatorname{Pr}_{#1}[#2]}

\newcommand{\ex}[1]{\mathbb{E}\left[#1\right]}
\newcommand{\exsub}[2]{\mathbb{E}_{#1}\left[#2\right]}
\newcommand{\exsubpar}[2]{\mathbb{E}_{#1}[#2]}

\newcommand{\expar}[1]{\mathbb{E}[#1]}
\newcommand{\eps}{\epsilon}
\newcommand{\opt}{\mathrm{OPT}}
\newcommand{\ind}{\mathds{1}}
\newcommand{\PP}{\mathcal{P}}
\newcommand{\Hist}{\mathcal H}
\newcommand{\dd}{\mathrm d}
\newcommand{\policyct}{P_{\mathrm{CT}}}
\newcommand{\policyunb}{P_{\mathrm{CT}\infty}}
\newcommand{\policydiv}{P_{\mathrm{CT}\mathbb{R}}}
\newcommand{\convas}{\xrightarrow[u\to\infty]{\mathrm{a.s.}}}

\title{\boldmath\textbf{Improved Approximation Guarantees for\\Periodic-Review Stochastic Inventory Control}}
\author{Danny Segev\thanks{School of Mathematical Sciences, Tel Aviv University,
Tel Aviv 69978, Israel. Email: \texttt{segevdanny@tauex.tau.ac.il}. Research supported by Israel Science Foundation grant 767/26.}}
\date{}

\begin{document}

\maketitle
\thispagestyle{empty}

\begin{abstract}
\noindent In periodic-review stochastic inventory control, replenishment policies must balance holding and backlogging
costs while responding to uncertain demands and evolving forecasts that may be
arbitrarily correlated and nonstationary. In this general information
setting, Levi, P{\'a}l, Roundy, and Shmoys (\textit{Mathematics of Operations Research} 2007) established that the fully adaptive optimum can be approximated within factor $2$, which had remained the best-known performance guarantee since then. Their marginal cost accounting and
dual-balancing framework made this comparison possible without computing
optimal decisions across future trajectories, opening a broad research
program on approximation algorithms for stochastic inventory systems.

\medskip \noindent We break this long-standing barrier for periodic-review stochastic inventory control, showing how to compete against the fully adaptive benchmark within factor $\frac{ e }{ e-1 } \approx 1.582$ while retaining general demand correlations and evolving forecasts. We attain these advances by discovering a structural connection between
marginal inventory accounting and weighted stopping problems, where each unit's conditional holding
charge forms a  supermartingale, while postponing its order
incurs a history-dependent backlog charge. To leverage the random thresholding approach of Ma and Tzamos (\textit{ICML} 2023) in this context, we develop a potential-function analysis, leading to an $\frac{ e }{ e-1 }$-approximation with respect to any unit's stopping problem. Subsequently, our common-threshold policy coordinates these marginal decisions into a global replenishment policy without additional loss.
\end{abstract}

\vspace{1em}

\newpage
\thispagestyle{empty}
\tableofcontents
\newpage

\pagenumbering{arabic}
\section{Introduction}\label{sec:intro}

Inventory decisions determine whether firms can meet demand when it arises and how much capital they must commit before knowing what customers will need. Across manufacturing, retail, and distribution, ordering too early ties up resources in stock, whereas ordering too late creates shortages and delays. The classic periodic-review stochastic inventory control model captures this recurring tradeoff through a sequence of replenishment decisions made as demand and new information arrive. These decisions become particularly consequential in the presence of lead times, as any order commits supply prior to knowing the demand it will serve, while later forecasts may substantially revise the original assessment that justified it. Therefore, optimal policies must simultaneously account for uncertain demand and for the information that future observations may reveal. 

These features make inventory control a central setting in adaptive optimization, with fundamental questions that have remained compelling for decades. When should a firm commit to replenishment, and when should it wait for additional demand information? How should lead times and correlations across future demands influence the timing and size of its orders? Can a tractable policy account for the information that future observations may reveal without optimizing over every possible sequence of these observations? How closely can a computationally efficient replenishment policy match the best policy that responds to the entire evolving history? Due to the breadth and depth of previous literature in this context, the upcoming discussion focuses only on the specific topics that are most directly related to our work. For broader treatments of stochastic inventory models, we refer
interested readers to textbooks such as those of \citet[Chap.~9]{zipkin2000},
\citet[Chap.~7]{porteus2002}, \citet[Chap.~1]{sethi2005}, and
\citet[Chaps.~5--6]{axsater2015}. 

\paragraph{The cost-balancing program.} \citet{lprs2007} established that strong approximation guarantees are possible for periodic-review stochastic inventory control in very general information settings. Their dual-balancing policy attains expected cost at most twice that of an optimal adaptive policy, while allowing nonstationary correlated demands and evolving forecasts. In a nutshell, their methodological insight was to assign future holding costs to currently placed orders, which are balanced against the consequences of insufficient supply. This accounting scheme allows comparing tractable policies with the optimal fully adaptive policy without generating the latter's decisions across all possible future histories. These advances opened a broad research program in which related balancing ideas support provably good approximation guarantees for capacitated systems \citep{capacitated2008}, stochastic lot sizing with setup costs and positive lead times \citep{lot2013}, and serial multiechelon systems \citep{multiechelon2017}, to name a few applications. Such examples are part of a wider literature that develops and refines marginal-cost accounting and balancing ideas for lost sales, remanufacturing, and perishable inventory, including models with additional replenishment constraints; see, e.g.,
\citet{lost2008}, \citet{taozhou2014}, \citet{perishable2015},
\citet{perishablesetup2016}, \citet{capacitatedperishable2018}, and \citet{zhangayerwhite2023}. 

\paragraph{Informal contributions.}
In this paper, we break the long-standing $2$-approximation barrier for periodic-review stochastic inventory control, showing how to compete against the fully adaptive benchmark within factor $\frac{ e }{ e-1 } \approx 1.582$ while retaining general demand correlations and evolving forecasts. Our approach creates a structural bridge between inventory
accounting and stopping under uncertainty, arguing that for each unit label, its conditional expected holding charges form a nonnegative supermartingale, while each postponement incurs an observed history-dependent backlog charge.
This connection turns replenishment decisions into an interrelated collection of weighted stopping problems, for which we establish an $\frac{ e }{ e-1 }$-approximation by combining the random thresholding technique of \citet{matzamos2023} with potential-function arguments. 
To translate these separate per-unit guarantees into a complete inventory policy, we coordinate all unit decisions through a common-threshold policy, yielding feasible orders without additional loss.

\paragraph{Outline.} Section~\ref{subsec:model} provides a formal mathematical description of periodic-review stochastic inventory control, its fully adaptive benchmark, and our oracle access assumptions. Section~\ref{subsec:related} reviews  directly related work and summarizes central motivating questions. Section~\ref{subsec:contributions} presents our main results, along with a succinct overview of their underlying algorithmic developments and analytical ideas. 

\subsection{Model formulation}\label{subsec:model}

\paragraph{Demand and observations.} Given a single item, our objective is to meet its stochastic demand across $T$ discrete periods, indexed by $1,\ldots,T$. We make use of the integer-valued random variable $D_t \geq 0$ to designate the item's demand in each period $t 
\in [T]$. From an information perspective, at the beginning of this period, all past demands $D_1,\ldots,D_{t-1}$ are known, as well as all previously received external signals $\zeta_1,\ldots, \zeta_t$, such as advance demand information or revised forecasts.  We represent this cumulative knowledge by $\Hist_t = \sigma(D_1,\ldots,D_{t-1} ,\zeta_1, \ldots, \zeta_t)$, where $\sigma(\cdot)$ denotes the sigma-field generated by the listed random variables. Setting $\Hist_{T+1}=\sigma(\Hist_T,D_T)$, we obtain the filtration $\Hist_1\subseteq\Hist_2\subseteq\cdots\subseteq\Hist_{T+1}$, which records how information accumulates over time. Moving forward, saying that a quantity is $\Hist_t$-measurable means that its value depends only on the demands and signals observed by the beginning of period $t$. As explained below, this view allows us to consider general forecast updates without specifying a state-space model or a particular forecast-update equation, as long as appropriate oracles are accessible for policy implementation. It is important to emphasize that the demand and signal processes are exogenous, meaning that our ordering decisions affect neither their joint distribution nor which signals are observed. Demands and signals may otherwise be arbitrarily correlated and nonstationary.

\paragraph{Replenishment policies.}
At the beginning of period $t$, subsequently to observing the available history $\Hist_t$ yet prior to observing the current demand $D_t$, a replenishment policy $P$ should decide on an integer ordering quantity $Q_t^P \geq0$, possibly by utilizing an independent random seed $\xi_P$. For simplicity of presentation, our main text treats integer demands and ordering quantities; Appendix~\ref{app:divisible} will extend the policy we propose and its approximation guarantee to divisible quantities and nonnegative real-valued demand. An order placed at time $t$ arrives at the beginning of  period $t+L$, where the lead time $L$ is a known nonnegative integer. Letting $n=T-L$ be the last period in which ordered units can still arrive within our planning horizon, we  assume that $n\geq1$. If $L\geq T$, orders cannot arrive in time and it is optimal to avoid placing any orders. Now, upon order arrival, inventory dynamics proceed as follows:
\begin{itemize}
    \item Arriving supply first reduces outstanding backlog; any surplus becomes on-hand inventory. 
    
    \item Demand is then served from the available inventory, reducing its level.
    
    \item Any demand that cannot be immediately  met is added to the backlog and carried forward until sufficient supply arrives.
\end{itemize}
In terms of additional features, there are no  fixed ordering costs, whose inclusion leads to the stochastic lot-sizing problem, which is outside the scope of our work. On the other hand, even though we assume zero per-unit purchase costs, no initial inventory, and an empty initial pipeline, such features can be incorporated by employing standard ideas, as explained in Section~\ref{subsec:contributions}.

\paragraph{Inventory levels and cost function.} Based on the preceding discussion, the cumulative demand through period $t$ can be written as $D_{[t]} = \sum_{r \leq t} D_r$, whereas the cumulative order quantities of a given policy $P$ will be designated by  $Q_{[t]}^P = \sum_{s \leq t} Q_s^P$, with the convention that $D_{[0]}=Q_{[0]}^P=0$. To interpret these coordinates, note that precisely the orders placed in periods $1, \ldots, t-L$ have arrived before the demand of period $t$. Consequently, our net inventory after period-$t$ demand has occurred is $I_t^P = Q_{[t-L]}^P-D_{[t]}$. A positive balance translates to on-hand inventory, while a negative balance is outstanding backlog.

For each period $t \in [T]$, let $h_t\geq0$ and $p_t\geq0$ be deterministic per-unit holding and backlogging cost rates. Knowing these parameters in advance, at the end of period $t$, after scheduled replenishments have arrived and demand $D_t$ has occurred, any policy $P$ incurs a holding and backlogging cost of $h_t \cdot [I_t^P]^+ + p_t \cdot [-I_t^P]^+$. Here, only the term corresponding to the sign of net inventory actually contributes. As such, the random total cost of this policy across the entire planning horizon is given by
\begin{equation} \label{eqn:total_random_cost}
F(P)~~=~~\sum_{t \in [T]} \left( h_t \cdot [I_t^P]^+ + p_t \cdot [-I_t^P]^+ \right).    
\end{equation}

\paragraph{The fully adaptive benchmark.} We say that a policy $P$ is nonanticipatory, or adapted to the available information, when each ordering decision depends only on the current observed history $\Hist_t$ and its independent random seed $\xi_P$. Formally, \(Q_t^P\) must be measurable with respect to $\Hist_t \vee \sigma(\xi_P)$. Letting $\PP$ denote the family of computationally unbounded nonanticipatory policies  for the periodic-review model under consideration, we wish to devise efficiently implementable policies for competing against the infimum expected cost,
\begin{equation} \label{eqn:benchmark}
\opt~~=~~\inf_{P\in\PP}\exsub{\omega,\xi_P}{F(P)}.
\end{equation}
This expectation is taken over the exogenous realization \(\omega\) of demands and signals, as well as over the random seed \(\xi_P\), when $P$ is randomized. In other words, the comparison policies know the joint law of demand and signals, including how forecasts will be generated. However, its ordering decision at any time $t$ must still utilize only the available history $\Hist_t$ and its independent seed, but not future observations. As a side note, the infimum in definition~\eqref{eqn:benchmark} avoids the need for additional assumptions that would ensure existence of an optimal policy.

\paragraph{Bounded demand.} For ease of exposition, our main results assume a known finite integer bound $M$ on the total demand, with $D_{[T]}\leq M$ almost surely. Under this assumption, cumulative orders may be restricted to $Q_{[n]}^P \leq M$ without loss of generality, since ordering additional units cannot reduce backlog costs and can only add holding costs. While bounded-support demand formulations have been widespread in stochastic inventory control and lot sizing problems, we nevertheless dedicate Appendix~\ref{app:unbounded_demand} to unbounded demand, under the assumptions $\exsubpar{\omega}{D_{[T]}}<\infty$ and $h_T>0$. \citet{lprs2007} likewise require total demand with finite expectation, whereas the technical condition $h_T>0$ is implicitly being utilized in their work, to argue about divergence of expected holding costs as order quantities increase.

\paragraph{Conditional probability oracle.} To deal with arbitrarily structured demand and signal processes, we assume oracle access to cumulative-demand probabilities conditional on observed histories. Specifically, at the beginning of period $t$, our oracle takes as input a history index $s\in\{1,\ldots,t\}$, a target period $r\in\{s,\ldots,T\}$, and  threshold $u\in [M]$. It returns the realized value of $H_{s,r}(u) = \prsubpar{\omega}{D_{[r]}<u|\Hist_s}$, namely, the probability that cumulative demand through period $r$ is strictly below $u$, given the information available at the beginning of period $s$. As such, only histories actually observed during the policy's execution need to be supported. Still, these histories should remain accessible, and a query issued at time $t$ can ask about the conditional law associated with $\Hist_s$, for an earlier period $s$. Along with standard computational tracking, our running time bounds will count these calls as single operations, under exact arithmetic and comparisons; implementing an oracle of this form for a particular demand model is of course a separate question. 

Our oracle assumption can be compared with the computational requirements of \citet{lprs2007}. For an efficient implementation, their dual-balancing policy evaluates, for a proposed order quantity, its expected holding cost over the remaining horizon and the expected backlog cost in the arrival period of that order. More precisely, at any ordering period $t \in [n]$, given the observed history $\Hist_t$, the realized cumulative quantity $Q_{[t-1]}$, and a candidate order quantity $q\geq0$, their implementation requires evaluating:
\begin{itemize}
    \item $\exsubpar{\omega}{
    [Q_{[t-1]}-D_{[r]}+q]^+-[Q_{[t-1]}-D_{[r]}]^+
    | \Hist_t}$, for every $t+L \leq r \leq T$.

    \item $\exsubpar{\omega}{
    [D_{[t+L]}-Q_{[t-1]}-q]^+
   | \Hist_t}$.
\end{itemize}
Our oracle provides conditional cumulative-demand probabilities, which contain less information than the full conditional joint distribution. However, their expected-cost functions also depend only on cumulative-demand marginals. Thus, the two approaches incorporate related distributional information but specify different access requirements, and we do not claim an unconditional ordering of these computational assumptions. 

\subsection{Directly related work and motivating questions}\label{subsec:related}

To provide problem-specific context around our work, we proceed by reviewing three research threads in stochastic inventory control: marginal-cost accounting and balancing policies, single-unit decompositions, and approximation schemes under alternative demand representations. This overview will shed light on the core obstacles that should be resolved toward attaining improved  approximation guarantees while addressing general demand
correlations and evolving forecasts.

\paragraph{Marginal accounting and dual balancing.} As its starting point, our work builds on the marginal accounting scheme of \citet{lprs2007}, which assigns future holding costs to currently placed orders, balanced against the consequences of insufficient supply. This scheme makes the outcomes of ordering comparable
with those of waiting, without knowing optimal decisions at future histories.
Their dual-balancing policy employed such comparisons to define a replenishment policy whose expected total cost is within factor $2$ of optimal.
Subsequent work leveraged these accounting and balancing principles to tackle additional inventory constraints. Prominent examples include  capacitated systems, for which \citet{capacitated2008} introduced
forced marginal backlogging costs to capture shortages that
an ordering decision makes unavoidable in later periods,
obtaining a factor-$2$ guarantee.
\citet{lot2013} attained a $3$-approximation for stochastic
lot sizing with setup costs and positive lead times, while
\citet{shi2014} established a factor-$4$ guarantee when setup
costs and capacity constraints are both present.
For serial multiechelon systems, \citet{multiechelon2017}
developed balancing policies with guarantees between $2$ and $3$,
depending on the assumptions in place.
This literature shows how marginal accounting can accommodate
interactions created by shared capacity, setup costs, and
movement between inventory locations.

\paragraph{Single-unit decompositions.}
A closely related perspective studies inventory decisions
through individual units and the demands they eventually serve.
\citet{unit2008} developed this approach for uncapacitated serial
multiechelon systems with Markov-modulated demand and
stochastic lead times, assuming that orders do not cross.
Their representation allows studying
the decisions associated with individual units.
In finite-horizon analysis, they  decomposed the optimal
cost of a surrogate model into single-unit optima, and then selected
compatible monotone unit policies to recover an optimal policy
for the original system.
This work belongs to a broader line of structural analyses
in multiechelon inventory control, including the classical
echelon decomposition of \citet{clark1960}, the
state-dependent policies for Markov-modulated demand of
\citet{chen2001}, and the approximate policies for
time-correlated demand due to \citet{dong2003}. Our work exploits the single-unit perspective to handle
two separate tasks: assigning costs to individual units
and ensuring that their approximate decisions can be implemented together. That said, unlike the above-mentioned papers, we work directly under the general  information structure of
Section~\ref{subsec:model}, without constructing or solving the associated state-space dynamic program. 

\paragraph{Demand representations and computational access.} Broadly speaking, the performance guarantees available for stochastic inventory
control depend on both the demand structure and its computational
representation. A well-known example is finite-horizon single-item inventory control with
independent discrete demands, for which \citet{halman2009} proposed a fully polynomial-time approximation
scheme. In regard to nonlinear newsvendor and stochastic
lot-sizing problems, \citet{halman2012} examined oracle representations of demand
and cost functions allowing approximation schemes under
specific assumptions, and established impossibility results for other
oracle formulations.
These results make the choice of computational access an
essential part of any comparison between approximation
guarantees. 

Our formulation accommodates evolving forecasts through
the conditional demand laws associated with observed histories,
without prescribing a particular forecast-update equation.  In this context,
\citet{heath1994} introduced the martingale model of forecast
evolution, while \citet{gallego2001} studied how advance demand
information enters replenishment decisions.
\citet{iida2006} developed approximate solutions for a dynamic
forecast-inventory model, and \citet{lu2006} derived inventory
policies and cost bounds in the presence of forecast
updates. A separate line of work studies access through demand samples, where \citet{sampling2007} established sample requirements for
near-optimal policies in newsvendor and multiperiod inventory
models with independent demands, and \citet{data2015} sharpened performance bounds for
the sample-average newsvendor solution. 

\paragraph{Motivating questions.} With two decades passing since the introduction of cost balancing policies into approximation frameworks for stochastic inventory control, our work is primarily motivated by the following questions:
\begin{itemize}
 \item \textit{Performance guarantees}: Can the long-standing $2$-approximation for periodic-review stochastic inventory control be surpassed while retaining arbitrary demand correlations and forecast updates?
 
 \item \emph{Analytical approach}: Can we exploit the shared structure of marginal holding and backlog costs to derive sharper comparisons with the fully adaptive benchmark?
    
 \item \emph{Coordination and implementation}: Can we coordinate approximate decisions for single inventory units into feasible  replenishments and efficiently implement the resulting policy?
\end{itemize}

\subsection{Main contributions and technical ideas} \label{subsec:contributions}

The main contribution of this work is to improve upon the $2$-approximation of \citet{lprs2007} for periodic-review stochastic inventory control, showing how to provably compete against the fully adaptive benchmark within factor $\frac{ e }{ e-1 } \approx 1.582$ while retaining general demand correlations and evolving forecasts. The following theorem formalizes the precise guarantees attained by our common-threshold policy $\policyct$, including efficient implementation under the conditional-probability oracle
of Section~\ref{subsec:model}.

\begin{theorem}\label{thm:main}
For bounded-demand periodic-review stochastic inventory control, the common-threshold policy $\policyct$ guarantees an expected total cost of $\exsubpar{\omega,\rho}{F(\policyct)} \leq \frac{e}{e-1} \cdot \opt$. This policy requires $O(T^3\log M)$ conditional cumulative-demand probability queries and arithmetic operations overall.
\end{theorem}

\paragraph{Connection to supermartingale stopping.} At a high level, our approach establishes a connection between marginal inventory
accounting and a weighted generalization of the supermartingale stopping problem, studied by
\citet{matzamos2023}.
In their basic setting, a decision maker observes stopping prices
that form a nonnegative supermartingale and pays a unit charge
whenever the decision to stop is postponed. For inventory control, taking the single-unit perspective, we consider an analogous weighted stopping problem in which 
waiting charges form an arbitrary nonnegative adapted process, while stopping prices form a
nonnegative  supermartingale with terminal value zero. Here, our charge for postponing a unit's
order depends on the current demand forecast and can vary across
both periods and observed histories. 
While the work of \citet{matzamos2023} can handle integer-valued information
costs by introducing unit-cost virtual stages, our waiting
charges involve conditional demand probabilities, taking arbitrary real values, even when demands and orders are integer-valued. In spite of these distinctions, we prove that by extending their accumulated-clock and random-threshold construction, its $\frac{ e }{ e-1 }$-approximation can be migrated to single-unit stochastic inventory control. To analyze arbitrary history-dependent waiting charges, we introduce a potential function that controls the expected cost remaining from any competing stopping time.
This guarantee is transferred to inventory control by arguing that the marginal holding charges form supermartingales and that the individual stopping decisions can be coordinated into feasible orders. In what follows, we elaborate on the guiding principles of this approach and its implementability.

\paragraph{Main idea 1: From inventory accounting to stopping.}
In Section~\ref{sec:accounting}, we index supply by cumulative unit labels and express the expected inventory cost as a sum of unitwise charges, plus an unavoidable backlog cost prior to the arrival of any order, similar to \citet{lprs2007}. For a fixed label, ``stopping'' means ordering that unit, and as such, its stopping price is the conditional expected holding cost from arrival through the end of the planning horizon. Its waiting charge is the conditional expected backlog cost made unavoidable by delaying that order for one period. Crucially, these processes depend only on the exogenous history and the label in question, meaning that every policy faces the same pair of charges. Along these lines, we prove that the stopping price is a supermartingale by showing that every period advancement removes a nonnegative term from the remaining holding-cost tail; then, averaging over forecast updates leads to the required conditional inequality.

\paragraph{Main idea 2: Potential-function analysis.}
In Section~\ref{sec:stopping}, extending the algorithmic approach of \citet{matzamos2023}, we accumulate the ratio of current waiting charge to current stopping price and decide to stop when this clock crosses an independent random threshold. To analyze this rule, we first average over the threshold, obtaining a charge for each stage that still depends on the observed process. Subsequently, we bound this charge in two ways: by $\frac{ e }{ e-1 }$ times the current waiting charge, and by the conditional expected decrease of an appropriately defined potential, depending on the current stopping price and the truncated clock. For any competing stopping rule, we charge the stages before it stops to its waiting costs, and from there  onward, we sum the potential inequalities to bound our remaining expected cost in terms of its stopping price.

\paragraph{Main idea 3: Coordinating unit decisions.}
In Section~\ref{sec:algorithm}, we tie between the weighted stopping guarantees of individual labels through a single threshold. Holding charges are nondecreasing in the label, whereas waiting charges are nonincreasing, meaning that the resulting clocks are nonincreasing. Consequently, the labels that have crossed our common threshold form a growing prefix over time, and its successive increases define feasible order quantities. We will show that the ordering time of each unit agrees with its weighted stopping time, implying that unitwise guarantees can be aggregated to attain an $\frac{ e }{ e-1 }$-approximation against the fully adaptive optimum. Additionally, this prefix monotonicity permits binary search for each new cumulative order quantity, leading to an $O(T^3\log M)$-time implementation. To establish tightness, we describe a family of single-period Bernoulli-demand instances on which our policy's approximation ratio approaches $\frac{ e }{ e-1 }$, thereby showing that its analysis is best possible. 

\paragraph{Model extensions.} It is worth pointing out that initial stock, backlog, and outstanding orders can be incorporated via known arguments based on the accounting scheme of \citet{lprs2007}, which separates committed supply and unavoidable costs from decisions about new purchases. Deterministic  per-unit purchase prices can also be handled through  standard ideas, as described in Section~4.6  of their work. By contrast, unbounded demand and divisible quantities require special treatment. For this purpose, Appendix~\ref{app:unbounded_demand} operates under two assumptions, similar to those of \citet{lprs2007}: (1)~Total demand has finite expectation, i.e., $\exsubpar{\omega}{D_{[T]}}<\infty$; and (2)~Inventory left when the planning horizon ends incurs a
strictly positive  holding cost, $h_T>0$. Here, we prove that an adapted common-threshold policy places finite orders almost surely, that its cumulative order quantities have finite expectations, and that our $\frac{ e }{ e-1 }$-approximation continues to hold. Appendix~\ref{app:divisible} allows real-valued demand and divisible orders under a known finite demand bound. In this context, we replace unit sums by integrals, verify feasibility of the continuous passing prefixes, and transfer our stopping comparison to the divisible benchmark, yielding an $\frac{ e }{ e-1 }$-approximation relative to the divisible-order optimum.

\section{Marginal Cost Accounting and the Stopping Reduction}\label{sec:accounting}

In what follows, building on the marginal-cost accounting scheme of \citet{lprs2007}, we develop the connection between inventory accounting and weighted
stopping by expressing
replenishment decisions through individual units, each facing a choice
between ordering now and incurring further backlog charges while waiting. For this purpose, in Section~\ref{subsec:labels}, we assign cumulative labels to supply and
demand, represent each unit's ordering time as a stopping time, and
express inventory and backlog by counting their corresponding labels.
Section~\ref{subsec:charges} then defines policy-independent holding and
waiting charges, proving that together with an unavoidable initial backlog cost, their expected sum recovers a policy's
expected total cost.
Finally, Section~\ref{subsec:supermart} establishes that each unit's
holding-charge process is a nonnegative supermartingale and records
the charges' monotonicity across labels, providing the properties
required for our stopping analysis 
and the subsequent coordination of unit decisions into feasible
replenishment orders.

\subsection{Cumulative supply labels}\label{subsec:labels}

\paragraph{Ordering times of individual units.} Let us recall that $D_{[t]}$ stands for cumulative demand through period $t$ and that $Q_{[t]}^P$ is cumulative supply ordered through this period by a given policy $P$. We say that demand for unit $u \in [M]$ has occurred by period $t$ precisely when $D_{[t]}\geq u$. Similarly, with respect to the policy $P$, unit $u$ has been ordered by period $t$ precisely when $Q_{[t]}^P \geq u$. Consequently, its ordering time can be written as 
\begin{equation} \label{eqn:stopping_policy_unit}
S_P(u) ~~=~~ \min \left\{t\in[n]:Q_{[t]}^P\geq u \right\},    
\end{equation}
with the minimum of the empty set defined as $n+1$. This last value symbolically denotes a unit that is never ordered within the useful planning horizon rather than an actual order in period $n+1$.

For a fixed label $u$ and any decision period $t \in [n]$, the event $\{S_P(u)\leq t\}$ is exactly $\{Q_{[t]}^P\geq u\}$. Under a deterministic policy, $Q_{[t]}^P$ is determined by the information $\Hist_t$ available at the beginning of period $t$, implying that $S_P(u)$ is a stopping time with respect to $(\Hist_t)$. For a randomized policy, the same statement holds with respect to the enlarged filtration,  $(\Hist_t\vee\sigma(\xi_P))$, which includes the policy's independent random seed $\xi_P$ in addition to the observed history. Moreover, $S_P(u)$ is nondecreasing in $u$, i.e., for any two labels $u\leq u'$, we have $S_P(u)\leq S_P(u')$ on every realization, since replenishment policies cannot order a higher-numbered cumulative unit before ordering all lower-numbered ones.

\paragraph{Net inventory in these coordinates.} Circling back to Section~\ref{subsec:model}, we remind readers that, due to the fixed lead time $L$, our end-of-period-$t$ net inventory is $I_t^P = Q_{[t-L]}^P-D_{[t]}$. Here, the cumulative supply received by period $t$ is the prefix $1,\ldots,Q_{[t-L]}^P$ of unit labels $1, \ldots, M$, whereas the demand that has occurred thus far is the prefix $1,\ldots,D_{[t]}$. Therefore, at any time, a unit is held exactly when it belongs to the first prefix and not the second; it is backlogged exactly when it belongs to the second prefix and not the first. Hence, the positive and negative parts of $I_t^P$ can be written as
\begin{align}
[I_t^P]^+ ~~&=~~ [Q_{[t-L]}^P-D_{[t]}]^+
~~=~~\sum_{u \in [M]}\ind\{D_{[t]}<u\} \cdot \ind\{S_P(u)\leq t-L\}, \label{eq:holdunits}\\
[-I_t^P]^+ ~~&=~~ [D_{[t]}-Q_{[t-L]}^P]^+
 ~~=~~\sum_{u \in [M]}\ind\{D_{[t]}\geq u\} \cdot \ind\{S_P(u)>t-L\}.  \label{eq:backunits}
\end{align}
The first sum counts labels that have already arrived but have not yet been demanded; the second counts labels that have already been demanded but have not yet arrived. For $t\leq L$, no new unit has arrived and every event $\{S_P(u)>t-L\}$ holds automatically.

\subsection{Ordering and backlogging charges}\label{subsec:charges}

\paragraph{The charging processes, $\alpha_t(u)$ and $\beta_t(u)$.} Focusing on a single unit $u \in [M]$, we observe that when it is ordered in period $t \leq n$, it arrives in period $t+L$ and incurs holding cost in every period \(r\geq t+L\) for which \(D_{[r]}<u\). Consequently, its conditional expected holding cost is 
\begin{equation}\label{eq:a}
 \alpha_t(u) ~~=~~ \exsub{\omega}{  \sum_{r=t+L}^T h_r \cdot \ind\{D_{[r]}<u\} \middle| \Hist_t}.
\end{equation}
We set $\alpha_{n+1}(u)=0$, consistently with the convention that an unordered unit does not create holding costs. On the other hand, when this unit is not ordered by period $t$, it is unavailable in period $t+L$, and thus incurs a backlog charge in that period when $D_{[t+L]}\geq u$. The conditional expectation of this charge is
\begin{equation}\label{eq:b}
 \beta_t(u) ~~=~~ p_{t+L} \cdot \prsub{\omega}{D_{[t+L]}\geq u \middle| \Hist_t}.
\end{equation}
The different time spans in these definitions are deliberate. Ordering now commits unit $u$ to arrive in period $t+L$, and $\alpha_t(u)$ includes all holding periods that can follow that arrival. Waiting  leaves the arrival decision open, and thus $\beta_t(u)$ includes only the next backlog charge (in period $t+L$), made unavoidable by this delay. Both $\alpha_t(u)$ and $\beta_t(u)$ are conditional expectations determined by the current history $\Hist_t$, even though the actual charges occur later.

Importantly, the processes $\alpha_t(u)$ and $\beta_t(u)$ are policy independent. Indeed, by inspecting definitions~\eqref{eq:a} and~\eqref{eq:b}, at the same exogenous history, every policy faces the same charges. As such, any policy only affects  whether and when it acquires this label, but not the future demand and signal law utilized to evaluate that acquisition. We can therefore compare different policies through their ordering times, based on a single pair of processes.

\paragraph{Expected cost decomposition.}
To express the expected  total cost of a given policy $P$ through these charging processes, we first account for the initial $L$ periods. Since the system starts without on-hand inventory or outstanding orders, supply cannot arrive before period $L+1$. Every policy therefore incurs the same random backlog cost $\sum_{t \in [L]} p_tD_{[t]}$, and we denote its expectation by $K_0=\exsubpar{\omega}{\sum_{t \in [L]} p_tD_{[t]}}$. The following lemma decomposes a policy's expected total cost into this unavoidable initial cost and the expected charges associated with individual units. An intuitive interpretation of the resulting expression is that each unit accumulates backlog charges $\beta_t(u)$ prior to its ordering time $S_P(u)$, at which point it is assigned the conditional holding-cost charge $\alpha_{S_P(u)}(u)$. In essence, this lemma reformulates the marginal-cost accounting of
\citet{lprs2007} as a sum of unitwise stopping costs, and we provide its detailed proof to make
the conditioning at random ordering times explicit.

\begin{lemma} \label{lem:unitcost}
For every replenishment policy $P$,
\[  \exsub{\omega,\xi_P}{F(P)} ~~=~~ K_0+\sum_{u \in [M]}\exsub{\omega,\xi_P}{\alpha_{S_P(u)}(u)        +\sum_{t<S_P(u)}\beta_t(u)}. \]
\end{lemma}

\paragraph{Per-unit holding and backlog costs.} To establish this result, we first consider the scenario where $P$ is a deterministic policy, and subsequently extend our argument to randomized policies. For this purpose, let us fix a unit $u$, recalling that $S_P(u) = \min \{t\in[n]:Q_{[t]}^P\geq u \}$ stands for its ordering time with respect to $P$. On the event $\{S_P(u)=t\}$, the unit's realized holding cost is $\sum_{r=t+L}^T h_r \cdot \ind\{D_{[r]}<u\}$, and there is no holding cost when $S_P(u)=n+1$. As such, the expected total holding cost attributable to unit $u$ is 
\begin{align}
&\exsub{\omega}{\sum_{t \in [n]}\ind\{S_P(u)=t\} \cdot \sum_{r=t+L}^T h_r \cdot \ind\{D_{[r]}<u\}} \nonumber \\
&\quad=~~\sum_{t \in [n]}\exsub{\omega}{\ind\{S_P(u)=t\} \cdot \exsub{\omega}{\sum_{r=t+L}^T h_r \cdot \ind\{D_{[r]}<u\} \middle| \Hist_t}} \label{eqn:lem_unitcost_1} \\
&\quad=~~\sum_{t \in [n]}\exsub{\omega}{\ind\{S_P(u)=t\} \cdot \alpha_t(u)} \label{eqn:lem_unitcost_2} \\
& \quad=~~\exsub{\omega}{\alpha_{S_P(u)}(u)}. \label{eqn:lem_unitcost_2.5}
\end{align}
Here, equality~\eqref{eqn:lem_unitcost_1} is obtained by utilizing the tower property, with the $\Hist_t$-measurable indicator $\ind\{S_P(u)=t\}$ taken outside the conditional expectation, while equality~\eqref{eqn:lem_unitcost_2} follows from the definition of $\alpha_t(u)$ in~\eqref{eq:a}.

For the same unit $u$, it has not arrived by the beginning of period $t+L$ if and only if $S_P(u)>t$. We therefore incur the backlog cost $p_{t+L}$ of that period precisely when both $D_{[t+L]}\geq u$ and $S_P(u)>t$. Consequently, the expected total backlog cost attributable to unit $u$ beyond the first $L$ periods can be written as
\begin{align}
&\exsub{\omega}{
    \sum_{t\in[n]} p_{t+L}
    \cdot\ind\{D_{[t+L]}\geq u\}
    \cdot\ind\{S_P(u)>t\}}
    \nonumber\\
&\quad=~~
    \sum_{t\in[n]}\exsub{\omega}{
    \ind\{S_P(u)>t\}\cdot p_{t+L}
    \cdot\prsub{\omega}{
        D_{[t+L]}\geq u\middle|\Hist_t}}
    \label{eqn:lem_unitcost_3}\\
&\quad=~~
    \sum_{t\in[n]}\exsub{\omega}{
    \ind\{S_P(u)>t\}\cdot\beta_t(u)}
    \label{eqn:lem_unitcost_4}\\
&\quad=~~
    \exsub{\omega}{
    \sum_{t < S_P(u)}\beta_t(u)}. \label{eqn:lem_unitcost_45}
\end{align}
As before, equality~\eqref{eqn:lem_unitcost_3} follows from the tower property, since the $\Hist_t$-measurability of $\ind\{S_P(u)>t\}$ allows taking this indicator outside the conditional expectation, and equality~\eqref{eqn:lem_unitcost_4} substitutes the definition of $\beta_t(u)$ in~\eqref{eq:b}. 

\paragraph{Cost aggregation.} As shown in equation~\eqref{eqn:total_random_cost}, we know that the expected total cost $\exsubpar{\omega}{F(P)}$ of the policy $P$ can be written as $\exsubpar{\omega}{\sum_{t \in [T]} ( h_t \cdot [I_t^P]^+ + p_t \cdot [-I_t^P]^+ )}$. To recover the claimed representation of Lemma~\ref{lem:unitcost}, starting from the first term, the expected total holding cost of $P$ is given by
\begin{align}
&\exsub{\omega}{
    \sum_{t \in [T]} h_t \cdot [I_t^P]^+} \nonumber \\
&\quad=~~
    \sum_{u\in[M]} \exsub{\omega}{
    \sum_{t=L+1}^T
    h_t\cdot\ind\{D_{[t]}<u\}
    \cdot\ind\{S_P(u)\leq t-L\}} \label{eqn:lem_unitcost_6} \\
&\quad=~~
    \sum_{u\in[M]}\exsub{\omega}{
    \sum_{r\in[n]}\ind\{S_P(u)=r\}
    \cdot\sum_{t=r+L}^T
    h_t\cdot\ind\{D_{[t]}<u\}} \label{eqn:lem_unitcost_7}\\
&\quad=~~
    \sum_{u\in[M]}
    \exsub{\omega}{\alpha_{S_P(u)}(u)}. \label{eqn:lem_unitcost_8}
\end{align}
Here, equality~\eqref{eqn:lem_unitcost_6} is obtained by plugging in~\eqref{eq:holdunits}, interchanging the order of summation, and removing the initial periods $1, \ldots, L$, where no holding cost is incurred. To arrive at equality~\eqref{eqn:lem_unitcost_7}, we write $\ind\{S_P(u)\leq t-L\} = \sum_{r \in [t-L]} \ind\{S_P(u) = r\}$ and again interchange the order of summation. The final equality~\eqref{eqn:lem_unitcost_8} applies the fixed-unit holding-cost calculation~\eqref{eqn:lem_unitcost_2.5}, including the convention $\alpha_{n+1}(u)=0$.

Similarly, the second term, representing the policy's expected total backlog cost, can be expressed as
\begin{align}
&\exsub{\omega}{
    \sum_{t \in [T]} p_t \cdot [-I_t^P]^+} \nonumber \\
&\quad=~~ \exsub{\omega}{\sum_{t \in [L]} p_t D_{[t]}} +\exsub{\omega}{
    \sum_{t=L+1}^T p_t
    \cdot[D_{[t]}-Q_{[t-L]}^P]^+} \label{eqn:lem_unitcost_9}\\
&\quad=~~
    K_0+ \sum_{u\in[M]} \exsub{\omega}{
    \sum_{t=L+1}^T
    p_t\cdot\ind\{D_{[t]}\geq u\}
    \cdot\ind\{S_P(u)>t-L\}} \label{eqn:lem_unitcost_10} \\
&\quad=~~
    K_0+\sum_{u\in[M]}\exsub{\omega}{
    \sum_{t\in[n]}
    p_{t+L}\cdot\ind\{D_{[t+L]}\geq u\}
    \cdot\ind\{S_P(u)>t\}} \nonumber\\
&\quad=~~
    K_0+\sum_{u\in[M]}
    \exsub{\omega}{
    \sum_{t < S_P(u)}\beta_t(u)}. \label{eqn:lem_unitcost_11}
\end{align}
In this case, equality~\eqref{eqn:lem_unitcost_9} substitutes $[-I_t^P]^+ = [D_{[t]}-Q_{[t-L]}^P]^+$ and splits the planning horizon into periods $1, \ldots, L$ and $L+1, \ldots, T$. Equality~\eqref{eqn:lem_unitcost_10} is obtained by recalling that $K_0 = \exsubpar{\omega}{\sum_{t \in [L]} p_t D_{[t]}}$, along with plugging in~\eqref{eq:backunits} and interchanging the order of summation. Finally, equality~\eqref{eqn:lem_unitcost_11} applies the fixed-unit backlog-cost calculation~\eqref{eqn:lem_unitcost_45}.

\paragraph{Randomized policies.}
Now, suppose that the policy $P$ is randomized, and let $\xi_P$ be its random seed, independent of the entire exogenous demand and signal process. At each period $t$, its ordering decisions are determined by the observed history together with this seed. Thus, the events $\{S_P(u)=t\}$ and $\{S_P(u)>t\}$ belong to the enlarged sigma-field $\Hist_t\vee\sigma(\xi_P)$. Our independence assumption ensures that including this seed does not change conditional expectations of quantities determined by the exogenous process, meaning that $\exsubpar{\omega,\xi_P}{ Z | \Hist_t\vee\sigma(\xi_P)}
= \exsubpar{\omega}{Z|\Hist_t}$ almost surely, for every integrable random variable $Z=Z(\omega)$.
Applying this identity to the holding and backlog quantities in~\eqref{eq:a} and~\eqref{eq:b} shows that their conditional expectations remain $\alpha_t(u)$ and $\beta_t(u)$ under the enlarged filtration. Consequently, the conditioning arguments in the fixed-unit calculations remain valid, since the  indicators $\ind \{S_P(u)=t\}$ and $\ind \{S_P(u)>t\}$ are measurable with respect to the enlarged information, and the conditional charges are unchanged. Our subsequent aggregation utilizes only finite sums and linearity of expectation; repeating these steps with expectations over both $\omega$ and $\xi_P$ therefore leads to 
\[  \exsub{\omega,\xi_P}{F(P)} ~~=~~ K_0+\sum_{u \in [M]}\exsub{\omega,\xi_P}{\alpha_{S_P(u)}(u)        +\sum_{t<S_P(u)}\beta_t(u)}, \]
which establishes Lemma~\ref{lem:unitcost} for randomized policies as well.

\subsection{The supermartingale property}\label{subsec:supermart}

\paragraph{Background.}
A process $(X_t)$ is called adapted if its value at each time $t$ can be determined from the observed history $\Hist_t$. Assuming that $\exsub{\omega}{|X_t|}<\infty$ for every $t$, it is a supermartingale if, given the current history, its expected next value is no larger than its current value, i.e., $\exsubpar{\omega}{X_{t+1}|\Hist_t} \leq X_t$ almost surely. Repeatedly applying this inequality  through the tower property leads to the same comparison with any fixed future period $s > t$, namely, $\exsub{\omega}{X_s|\Hist_t}\leq X_t$ almost surely.
In particular, $\exsub{\omega}{X_s}\leq\exsub{\omega}{X_t}$, implying that the process's expected value is nonincreasing over time. For additional background on conditional expectation and the basic properties
of martingales and supermartingales, we refer readers to
\citet[Secs.~4.1--4.2]{durrett2019}.

\paragraph{The holding-cost process is a supermartingale.}
We now show that, for each fixed label $u$, its process of conditional expected holding charges $(\alpha_t(u))$ forms a  supermartingale over time. Specifically, our proof will argue that the expected decrease $\alpha_t(u)
-\exsub{\omega}{\alpha_{t+1}(u)\middle|\Hist_t}$ equals the conditional expected holding cost incurred in period $t+L$ if ordered now. This property will be crucial to employing the weighted  stopping result of Section~\ref{sec:stopping} and to comparing the expected cost of our proposed policy with that of any admissible stopping rule.

\begin{lemma}\label{lem:supermart}
$(\alpha_t(u))_{t \in [n+1]}$ is a nonnegative supermartingale with respect to $(\Hist_t)$, for each fixed label $u$. Moreover, for every period $t\in[n]$,
\[
\alpha_t(u)
-\exsub{\omega}{\alpha_{t+1}(u)\middle|\Hist_t}
~~=~~
h_{t+L}\cdot
\prsub{\omega}{D_{[t+L]}<u\middle|\Hist_t}
\qquad\text{almost surely}.
\]
\end{lemma}
\begin{proof}
We first check the basic requirements in the definition of a supermartingale. To this end, recalling that $\alpha_t(u) = \exsubpar{\omega}{  \sum_{r=t+L}^T h_r \cdot \ind\{D_{[r]}<u\} | \Hist_t}$ by definition~\eqref{eq:a}, we know that $\alpha_t(u)$ is a conditional expectation given $\Hist_t$, so its value is determined by the information available at time $t$. Thus, this process is adapted. Moreover, since all holding-cost coefficients are nonnegative, $0\leq \alpha_t(u)\leq \sum_{r=t+L}^T h_r$ almost surely, for every $t\in[n]$. The terminal value $\alpha_{n+1}(u)=0$ is also measurable. These bounds ensure that  every $\alpha_t(u)$ is integrable.

We next compare the charge $\alpha_t(u)$ at time $t$ with the expected charge $\exsubpar{\omega}{\alpha_{t+1}(u)|\Hist_t}$ at time $t+1$, based only on the information available at time $t$. For $t<n$, an order placed in period $t+1$ arrives in period $t+L+1$, meaning that its holding-cost tail starts one period later. Therefore, we almost surely have
\begin{align}
&\exsub{\omega}{\alpha_{t+1}(u)|\Hist_t} \nonumber \\
&\quad=~~
\exsub{\omega}{
    \exsub{\omega}{
        \sum_{r=t+L+1}^T
        h_r\cdot\ind\{D_{[r]}<u\}
        \middle|\Hist_{t+1}}
    \middle|\Hist_t} \nonumber \\
&\quad=~~
\exsub{\omega}{
    \sum_{r=t+L+1}^T
    h_r\cdot\ind\{D_{[r]}<u\}
    \middle|\Hist_t} \label{eqn:lem_supermart_1}\\
&\quad=~~
\exsub{\omega}{
    \sum_{r=t+L}^T
    h_r\cdot\ind\{D_{[r]}<u\}
    -h_{t+L}\cdot\ind\{D_{[t+L]}<u\}
    \middle|\Hist_t} \nonumber\\
&\quad=~~
\alpha_t(u)
-h_{t+L}\cdot
\prsub{\omega}{D_{[t+L]}<u\middle|\Hist_t}. \label{eqn:lem_supermart_2}
\end{align}
Here, equality~\eqref{eqn:lem_supermart_1} holds since $\Hist_t\subseteq\Hist_{t+1}$, and averaging the updated conditional expectation over the additional observations equals its conditional expectation based on the current history. Equality~\eqref{eqn:lem_supermart_2} is obtained by plugging in the definition of $\alpha_t(u)$. It remains to examine the last decision period, $t=n$. In this case, since the terminal value $\alpha_{n+1}(u)$ is zero, we obtain the claimed identity by observing that
\[
\alpha_n(u)
-\exsub{\omega}{\alpha_{n+1}(u)|\Hist_n}
~~=~~
\alpha_n(u)
~~=~~ \exsub{\omega}{  h_T \cdot \ind\{D_{[T]}<u\} \middle| \Hist_n} ~~=~~
h_T\cdot\prsub{\omega}{D_{[T]}<u\middle|\Hist_n}.
\]
In summary, $\alpha_t(u)
-\exsubpar{\omega}{\alpha_{t+1}(u)|\Hist_t}
=
h_{t+L}\cdot
\prsubpar{\omega}{D_{[t+L]}<u|\Hist_t}$ almost surely for every $t\in[n]$. Since the right-hand side is nonnegative, we conclude that $\exsub{\omega}{\alpha_{t+1}(u)|\Hist_t}\leq\alpha_t(u)$ almost surely. Together with the adaptedness and integrability verified above, it follows that $(\alpha_t(u))_{t \in [n+1]}$ is a nonnegative supermartingale.
\end{proof}

\paragraph{Monotonicity.} For our subsequent analysis, it is worth mentioning that $\alpha_t(u)$ is nondecreasing in $u$, whereas $\beta_t(u)$ is nonincreasing in $u$. Indeed, for every pair of labels $u\leq u'$, on every demand path, $\ind\{D_{[r]}<u\}\leq\ind\{D_{[r]}<u'\}$ and $\ind\{D_{[r]}\geq u\}\geq\ind\{D_{[r]}\geq u'\}$. Based on how $\alpha_t(\cdot)$ and $\beta_t(\cdot)$ are respectively defined in~\eqref{eq:a} and~\eqref{eq:b}, by taking conditional expectations with nonnegative coefficients, we have $\alpha_t(u)\leq \alpha_t(u')$ and $\beta_t(u)\geq \beta_t(u')$, as required. 
\section{The Weighted Supermartingale Stopping Theorem}\label{sec:stopping}

In what follows, we study a weighted stopping problem of the type arising from our unitwise cost decomposition, in which each waiting decision  incurs an observed charge, while stopping prices form a nonnegative supermartingale. Building on the accumulated clock and random-threshold rule of \citet{matzamos2023}, we propose a potential-function analysis that transfers their $\frac{e}{e-1}$-approximation from unit waiting charges to arbitrary nonnegative charges. Toward this objective, Section~\ref{subsec:stopobjective} formalizes the specific stopping objective we consider and explains how its prices and waiting charges correspond to our marginal-cost accounting reformulation. Section~\ref{subsec:stoprule} presents the randomized stopping rule and states its approximation guarantee. Section~\ref{subsec:stopoverview} provides a technical overview of our analysis, with its finer details appearing in Sections~\ref{subsec:stop-stage-proof}--\ref{subsec:stop-tail-proof}.

\subsection{The weighted stopping problem} \label{subsec:stopobjective}

\paragraph{Prices and waiting charges.} The random process we consider evolves along a sequence of $n$ stages, at which we may either stop or wait. At each stage $t$, stopping incurs a price $\alpha_t$, while waiting incurs a charge $\beta_t$ and allows us to make another decision at the next stage, after receiving any new information available then. Both $\alpha_t$ and $\beta_t$ are observed just before the decision at stage $t$. Thus, we know the immediate cost of either action, but generally do not know the prices and charges that will be observed in later stages. Here, a decision to wait may be useful, as stopping prices can fall later on or as the next observation can help us decide whether to continue waiting. If we wait at every decision stage, the process ends at a terminal stage $n+1$ with zero stopping price; in this case, we pay all $n$ waiting charges.

Formally, our problem of interest is defined on a probability space $(\Omega,\Hist,\mathbb P)$ with filtration $(\Hist_t)_{t \in [n+1]}$, where $\Hist_t$ represents all information available when the decision at stage $t$ is made. We assume that the stopping prices $\alpha_1,\ldots,\alpha_n,\alpha_{n+1}=0$ form an integrable nonnegative supermartingale with respect to this filtration. In particular, $\exsub{\mathbb P}{\alpha_{t+1}|\Hist_t}\leq\alpha_t$ almost surely, for every $t \in [n]$. It is worth noting that this property restricts the conditional expected price at the next stage; it does not require prices to decrease on every realization. Regarding the waiting charges $\beta_1, \ldots, \beta_n$, we assume that each $\beta_t$ is $\Hist_t$-measurable, with $\exsubpar{\mathbb P}{\sum_{t \in [n]}\beta_t}<\infty$. These  charges need not form a supermartingale and may depend on the observed history. The joint price and charge processes are exogenous, and their evolution is unaffected by our actions. 

\paragraph{Stopping rules.} A feasible stopping rule is a stopping time $S$ with respect to $(\Hist_t)$, taking values in $\{1,\ldots,n+1\}$. Similarly to Section~\ref{subsec:labels}, this requirement means that the decision to stop at stage $t$ is based only on the current history $\Hist_t$, without access to future observations. The cost of any such rule is given by $\mathcal C(S)=\sum_{t<S}\beta_t+\alpha_S$, which consists of the waiting charges paid at all stages preceding $S$, together with the stopping price at stage $S$. For example, stopping at stage $3$ costs $\beta_1+ \beta_2 + \alpha_3$, as our rule pays $\beta_1+ \beta_2$ for waiting at stages $1$ and $2$, but avoids $\beta_3$ due to stopping at stage $3$, where $\alpha_3$ is incurred. The virtual stage $n+1$ represents never stopping during the first $n$ stages, and its price is set to $\alpha_{n+1} = 0$. Our objective is to propose a feasible stopping rule whose expected cost $\exsub{\mathbb P}{\mathcal C(S)}$ is provably within a constant factor of the infimum such cost over all feasible stopping times.

\paragraph{The inventory interpretation.} To better understand how this stopping problem will connect to our unitwise cost decomposition, let us focus on a single unit  $u\in[M]$, and suppose that $\alpha_t=\alpha_t(u)$ and $\beta_t=\beta_t(u)$ are defined as in Section~\ref{subsec:charges}. Then, a stopping decision at stage $t\leq n$ corresponds to ordering this unit in period $t$, so that it arrives in period $t+L$. Its stopping price $\alpha_t$ is the conditional expected total holding cost attributable to this unit from its arrival through the end of our planning horizon. By contrast, waiting at stage $t$ means that this unit cannot arrive by period $t+L$. If cumulative demand has reached label $u$ by that period, we incur a backlog cost of $p_{t+L}$, and the waiting charge $\beta_t$ is the conditional expectation of this cost, given the information available at time $t$. Thus, each decision to wait accounts for one additional period of potential backlog, whereas stopping accounts for all subsequent holding costs.

For a replenishment policy $P$, the corresponding stopping time $S$ is precisely its ordering time $S_P(u)$, formally defined in Section~\ref{subsec:labels}. The stopping price $\alpha_S$ is the unit's conditional expected holding cost from its arrival through the end of the planning horizon, evaluated based on the information available when it is ordered. Similarly, the accumulated waiting charge $\sum_{t<S}\beta_t$ accounts for the backlog charges associated with all preceding delays, each evaluated based on the history available at that decision time. Importantly, the assumptions behind our weighted stopping problem are satisfied: Lemma~\ref{lem:supermart} establishes that the stopping prices $(\alpha_t)_{t\in[n+1]}$ form an integrable nonnegative supermartingale, while the waiting charges are nonnegative, adapted, and integrable by their definition. Moreover, Lemma~\ref{lem:unitcost} shows that summing the expected stopping costs over all labels and adding the unavoidable expected initial backlog cost $K_0$ recovers the expected total cost of $P$. When constructing a replenishment policy from per-unit stopping rules, however, we must  ensure that their stopping times are nondecreasing in the unit label, since a higher label cannot be ordered before a lower one. This condition guarantees that, at every decision period, the labels ordered so far form a prefix, consistently with the cumulative supply representation of Section~\ref{subsec:labels}.

\subsection{Stopping rule and main finding}\label{subsec:stoprule}

The upcoming discussion leverages the accumulated clock and random thresholding ideas of \citet{matzamos2023} to argue that we can still address the waiting charges and stopping prices arising in Section~\ref{subsec:stopobjective}. Even though their work is extendable to integer-valued waiting charges via virtual stages, it is important to emphasize that our current charges can take arbitrary real values due to involving conditional demand probabilities, even when demands and orders are integer-valued. As such, while the algorithmic ideas themselves nearly match, our analysis is  significantly different, operating through threshold-averaged stage charges and conditional potential inequalities.

\paragraph{The accumulated clock.}
For each stage $t \in [n]$, we make use of $r_t=\frac{\beta_t}{\alpha_t}$ to designate the ratio of waiting charge to stopping price at this stage, with the conventions $\frac{ 0 }{ 0 }=0$ and $\frac{ \beta }{ 0 } =\infty$ for $\beta>0$. For example, $r_t= \frac{ 1 }{ 4 }$ means that waiting at stage $t$ costs one quarter of the stopping price at that stage. Moving forward, we refer to $q_t=\sum_{s<t}r_s$ as the accumulated clock before incorporating stage $t$, thereby recording the relative cost of earlier postponements at the times when they were considered. This clock is clearly nondecreasing, even when the stopping prices are not. For convenience,  $z_t=q_t+r_t$ will capture its value after the current increment. 

\paragraph{The random threshold rule.} For a given threshold $v\in(0,1]$, our proposed rule stops at the first stage when the accumulated clock $(z_t)_{t \in [n]}$ reaches or exceeds $v$, namely, at stage $\tau(v)=\min\{t\leq n:z_t\geq v\}$. If there is no such crossing, we set $\tau(v)=n+1$, representing termination at the virtual stage after waiting through all $n$ decision stages. Technically speaking, at each stage $t$ actually reached, we first incorporate the increment $r_t$ to obtain $z_t$. If $z_t<v$, we wait and pay $\beta_t$; if $z_t\geq v$, we stop and pay $\alpha_t$, avoiding $\beta_t$. Thus, the current increment is incorporated before deciding whether to incur its associated waiting charge. Since $(z_t)_{t \in [n]}$ is a nondecreasing clock, $\{\tau(v)\leq t\}=\{z_t\geq v\}$ for every $t\in[n]$. The latter event belongs to $\Hist_t$, meaning that $\tau(v)$ is indeed a stopping time.

Now, rather than picking a fixed threshold, we draw a random threshold $\rho$ once, independently of the entire process, with density $f(v)=\frac{e^v}{e-1} \cdot \ind \{ v \in [0,1] \}$ and cumulative distribution function
\[ F_{\rho}(v) ~~=~~ 
\begin{cases}
0, \qquad & \text{when } v < 0 \\
\frac{e^v-1}{e-1} , \qquad & \text{when } v \in [0,1] \\
1, \qquad & \text{when } v > 1
\end{cases} \]
One way to sample $\rho$ is by drawing $U$ uniformly from $(0,1)$ and setting $\rho=\ln(1+(e-1)U)$. For each fixed threshold, the resulting rule is adapted to $(\Hist_t)$; with a random threshold, it is adapted to the filtration enlarged by this independent initial draw, $(\Hist_t \vee \sigma(\rho))$. From this point on, we use subscript $\mathbb P$ for expectations over the original process, subscript $\rho$ for expectations over the threshold alone, and subscript $(\mathbb P,\rho)$ when averaging over both sources of randomness.

\paragraph{Main result.} In the remainder of this section, we prove that our stopping rule achieves an $\frac{ e }{ e-1 }$-approximation against any stopping rule based on the available history.

\begin{theorem}\label{thm:stopping}
For every stopping time $S$ with respect to $(\Hist_t)$, 
\[
 \exsub{\mathbb P,\rho}{\mathcal C(\tau(\rho))}
 ~~\leq~~ \frac{e}{e-1} \cdot \exsub{\mathbb P}{\mathcal C(S)}.
\]
\end{theorem}

\subsection{Proof of Theorem~\ref{thm:stopping}: High-level overview} \label{subsec:stopoverview}

\paragraph{The truncated clock.} For purposes of analysis, let us introduce the truncated clock $\hat q_t=\min\{q_t,1\}$ for $t\in[n+1]$, and its increments $\delta_t=\hat q_{t+1}-\hat q_t$ for $t\in[n]$. This clock is nondecreasing, implying that $\delta_t\geq0$ even when the original clock becomes infinite. Moreover, for every $t\in[n]$, it is not difficult to verify that $\alpha_t\delta_t \leq \beta_t$ almost surely. 
In other words, the truncated increment $\delta_t$, valued at the current stopping price $\alpha_t$, is no larger than the current waiting charge $\beta_t$.

\paragraph{Expected cost and stage charges.} Let $G_t$ be the charge of our stopping rule at stage $t$, averaged over the random threshold $\rho$ alone, given by
\begin{align} 
G_t
 ~~&=~~
\exsub{\rho}{
 \alpha_t\cdot\ind\{\tau(\rho)=t\}
 +
 \beta_t\cdot\ind\{\tau(\rho)>t\}} \nonumber \\
 &=~~ 
 \alpha_t\cdot \prsub{\rho}{
 \tau(\rho)=t }  +  \beta_t\cdot \prsub{\rho}{ \tau(\rho)>t }.   \label{eqn:definition_Gt} 
\end{align}
Thus, $G_t$ remains a random variable on the original probability space $(\Omega,\Hist,\mathbb P)$, but no longer depends on the threshold draw. This definition includes the possibility of stopping earlier than stage $t$, in which case its charge at this stage is zero. To express the random cost $\mathcal C(\tau(\rho))$ of our stopping rule, note that we pay $\beta_t$ exactly at the stages preceding $\tau(\rho)$ and pay $\alpha_t$ if we stop at stage $t\leq n$; termination at stage $n+1$ adds no stopping cost. Consequently,
\[
 \mathcal C(\tau(\rho))
 ~~=~~
 \sum_{t \in [n]}
 \left(
 \alpha_t\cdot\ind\{\tau(\rho)=t\}
 +
 \beta_t\cdot\ind\{\tau(\rho)>t\}
 \right).
\]
Averaging first over the independent threshold $\rho$ and then over the original process, we conclude that the expected cost incurred is precisely
\begin{equation}\label{eq:stop-expected-cost}
\exsub{\mathbb P,\rho}{\mathcal C(\tau(\rho))}
 ~~=~~
 \exsub{\mathbb P}{
 \sum_{t \in [n]}
 \exsub{\rho}{
 \alpha_t\cdot\ind\{\tau(\rho)=t\}
 +
 \beta_t\cdot\ind\{\tau(\rho)>t\}}} ~~=~~
 \exsub{\mathbb P}{\sum_{t \in [n]}G_t}.
 \end{equation}

\paragraph{Upper-bounding $G_t$.}
The next lemma bounds our threshold-averaged charge $G_t$ by
$\frac{e}{e-1} \cdot \beta_t$, allowing us to compare this quantity with the waiting
cost incurred by competing stopping rules. Moreover, its sharper intermediate bound in terms of $\alpha_t\delta_t$
will support the potential argument discussed below.
The proof of this result is provided in Section~\ref{subsec:stop-stage-proof}, noting that its arguments rely on our specific threshold distribution but not on the supermartingale assumption.

\begin{lemma}
\label{lem:stop-stage-charge}
For every $t\in[n]$, the random variable $G_t$ is $\Hist_t$-measurable and satisfies
\[  0 ~~\leq~~ G_t ~~\leq~~ \frac{e}{e-1} \cdot \alpha_t \delta_t ~~\leq~~
 \frac{e}{e-1} \cdot \beta_t
 \qquad\text{almost surely}. \]
\end{lemma}

\paragraph{Stage potentials.}
To control the charges we incur from a competing stopping time onward, for every $t\in[n+1]$, let us  define
\begin{equation} \label{eqn:def_potential}
\Phi_t  ~~=~~  \frac{e}{e-1} \cdot (1-\hat q_t)\alpha_t,  
\end{equation}
with the interpretation that $1-\hat q_t$ measures the clock interval still available before hitting $1$, and $\alpha_t$ values this remaining length at the current stopping price. The process $(\Phi_t)_{t\in[n+1]}$ is nonnegative, adapted, and integrable. Indeed, adaptedness follows as both $\hat q_t$ and $\alpha_t$ are $\Hist_t$-measurable; nonnegativity and integrability follow from
$0\leq\Phi_t\leq\frac{e}{e-1} \cdot \alpha_t$. The next inequality, whose proof is given in Section~\ref{subsec:stop-potential-proof}, states that the conditional expected decrease in the potential $\Phi_t$ covers the threshold-averaged charge $G_t$. Its proof is the only part of our argument that relies on the supermartingale property of $(\alpha_t)$. 

\begin{lemma}
\label{lem:stop-potential}
$\Phi_t - \exsubpar{\mathbb P}{\Phi_{t+1}|\Hist_t} \geq G_t$ almost surely, for every $t\in[n]$.
\end{lemma}

\paragraph{Comparing from an adapted stopping time onward.} Our last auxiliary lemma, whose proof is presented in  Section~\ref{subsec:stop-tail-proof}, turns the single-stage potential inequality into a bound on the sum of charges from any competing stopping time $S$ onward. 

\begin{lemma}
\label{lem:stop-tail}
$\exsubpar{\mathbb P}{\sum_{t=S}^n G_t}
\leq \exsubpar{\mathbb P}{\Phi_S}$, for every stopping time $S$ with respect to $(\Hist_t)$.
\end{lemma}

\paragraph{Concluding the proof.} We now combine the preceding lemmas by splitting our
threshold-averaged charges at the competing rule's stopping time.
The charges before that time are bounded by its waiting costs,
while those from that time onward are bounded in expectation
by the potential. To formalize this idea, fix any stopping time $S$ with respect to $(\Hist_t)$ taking values in $\{1,\ldots,n+1\}$. Then,
\begin{align}
 \exsub{\mathbb P,\rho}{\mathcal C(\tau(\rho))}
 ~~&=~~
 \exsub{\mathbb P}{\sum_{t<S}G_t}
 +
 \exsub{\mathbb P}{\sum_{t=S}^n G_t} \nonumber\\
 &\leq~~
 \frac{e}{e-1} \cdot \exsub{\mathbb P}{\sum_{t<S}\beta_t}
 +
 \exsub{\mathbb P}{\Phi_S} \label{eqn:subsec_stoprule_1} \\
 &=~~
 \frac{e}{e-1} \cdot 
 \exsub{\mathbb P}{
 \sum_{t<S}\beta_t+(1-\hat q_S)\alpha_S} \label{eqn:subsec_stoprule_2}\\
 &\leq~~
 \frac{e}{e-1} \cdot 
 \exsub{\mathbb P}{\sum_{t<S}\beta_t+\alpha_S} \label{eqn:subsec_stoprule_3} \\
 &=~~
 \frac{e}{e-1} \cdot 
 \exsub{\mathbb P}{\mathcal C(S)}. \nonumber
\end{align}
Here, inequality~\eqref{eqn:subsec_stoprule_1} holds since $\sum_{t<S}G_t \leq \frac{ e }{ e-1 } \cdot \sum_{t<S}\beta_t$ almost surely by Lemma~\ref{lem:stop-stage-charge}, and since $\exsub{\mathbb P}{\sum_{t=S}^n G_t} \leq \exsub{\mathbb P}{\Phi_S}$ by Lemma~\ref{lem:stop-tail}. Equality~\eqref{eqn:subsec_stoprule_2} follows by substituting $\Phi_S =  \frac{e}{e-1} \cdot (1-\hat q_S)\alpha_S$. To obtain inequality~\eqref{eqn:subsec_stoprule_3}, we note that $0\leq \hat q_S\leq1$ and $\alpha_S\geq0$.

\paragraph{Randomized competing rules.} It is worth pointing out that this comparison also applies to a competing rule with a random
seed $\xi$ independent of all original information $\Hist_{n+1}$. Here, we enlarge the filtration to $(\Hist_t\vee\sigma(\xi))_{t\in[n+1]}$ and choose $\rho$
independently of both the original information and $\xi$.
Independence ensures that
\[
 \exsub{\mathbb P,\xi}{
 \alpha_{t+1}\middle|\Hist_t\vee\sigma(\xi)}
 ~~=~~
 \exsub{\mathbb P}{\alpha_{t+1}\middle|\Hist_t}
 ~~\leq~~ \alpha_t
 \qquad\text{almost surely}.
\]
Thus, the stopping prices remain a supermartingale, and the
same argument establishes Theorem~\ref{thm:stopping} with
$\exsub{\mathbb P,\xi}{\mathcal C(S)}$ on its right-hand side.
The left-hand side remains unchanged, since our stopping rule does not depend on $\xi$.

\subsection{Proof of Lemma~\ref{lem:stop-stage-charge}: Bounding the stage charge}\label{subsec:stop-stage-proof}

We first describe a useful representation of $G_t$ in terms of $\alpha_t$, $\beta_t$, and the cumulative distribution function $F_{\rho}$ of our random threshold $\rho$. To this end, since the accumulated clock $(z_t)_{t \in [n]}$ is nondecreasing, our stopping rule terminates at stage $t$ exactly when $z_{t-1} < \rho \leq z_t$, or equivalently, $q_t<\rho\leq q_{t+1}$; it waits at this stage exactly when $\rho> z_t = q_{t+1}$. Since the clock values $q_t$ and $q_{t+1}$  are independent of $\rho$,
\[
 \prsub{\rho}{\tau(\rho)=t}
 ~~=~~ \prsub{\rho}{q_t<\rho\leq q_{t+1}} ~~=~~
 F_\rho(\hat q_{t+1})-F_\rho(\hat q_t), \]
 and similarly,
 \[ \prsub{\rho}{\tau(\rho)>t} ~~=~~ \prsub{\rho}{\rho> q_{t+1}} ~~=~~
 1-F_\rho(\hat q_{t+1}).
\]
By substituting these probabilities into the definition of $G_t$ in equation~\eqref{eqn:definition_Gt}, we have
\begin{equation} \label{eqn:Gt_interms_Frho}
G_t ~~=~~ \alpha_t\cdot \left( F_\rho(\hat q_{t+1})-F_\rho(\hat q_t) \right)  +  \beta_t\cdot \left(  1-F_\rho(\hat q_{t+1})\right).    
\end{equation}

To upper-bound this random variable, we observe that
\begin{equation}\label{eq:stop-waiting-identity}
 \beta_t \cdot \left( 1-F_\rho(\hat q_{t+1}) \right) ~~ =~~
 \alpha_t\delta_t\cdot \left( 1-F_\rho(\hat q_{t+1}) \right).
\end{equation}
Indeed, if $\hat q_{t+1}=1$, both sides of~\eqref{eq:stop-waiting-identity} vanish, since $F_\rho(1)=1$.  If instead  $\hat q_{t+1}<1$, we know that no truncation has occurred, i.e., $\hat q_{t} = q_t$ and $\hat q_{t+1} = q_{t+1}$. When $\alpha_t>0$, this means that $\delta_t= \hat q_{t+1} - \hat q_{t} = q_{t+1} - q_t = r_t= \frac{ \beta_t }{ \alpha_t }$, and hence $\alpha_t\delta_t=\beta_t$. When $\alpha_t=0$, the condition $q_{t+1}<1$ forces $\beta_t=0$, since a positive waiting charge would imply $q_{t+1} \geq r_t = \frac{ \beta_t }{ 0 } = \infty$. Thus, $\alpha_t\delta_t=\beta_t$ in this scenario as well. Now, by plugging equation~\eqref{eq:stop-waiting-identity} back into representation~\eqref{eqn:Gt_interms_Frho}, it follows that
\begin{align}
G_t ~~&=~~ \alpha_t\cdot \left( F_\rho(\hat q_{t+1})-F_\rho(\hat q_t)  +  \delta_t\cdot \left(  1-F_\rho(\hat q_{t+1})\right) \right) \nonumber \\
&=~~ \frac{ \alpha_t }{ e-1 } \cdot \left( e^{ \hat q_{t+1} } - e^{ \hat q_t } +  \delta_t\cdot \left(  e - e^{ \hat q_{t+1}  } \right) \right) \label{eqn:stop-stage-proof_1} \\
&\leq~~ \frac{ e }{ e-1 } \cdot \alpha_t \delta_t  \label{eqn:stop-stage-proof_2} \\
&\leq~~ \frac{ e }{ e-1 } \cdot \beta_t . \label{eqn:stop-stage-proof_3}
\end{align}
Here, equality~\eqref{eqn:stop-stage-proof_1} holds since both $\hat q_t$ and $\hat q_{t+1}$ reside in $[0,1]$, and since $F_{\rho}(v) = \frac{e^v-1}{e-1}$ for all $v \in [0,1]$. Inequality~\eqref{eqn:stop-stage-proof_2} follows by noting that $e^{\hat q_{t+1} } - e^{ \hat q_t } = e^{ \hat q_{t+1} } \cdot (1- e^{ -\delta_t }) \leq \delta_t e^{ \hat q_{t+1} }$. Finally, inequality~\eqref{eqn:stop-stage-proof_3} is obtained by recalling that $\alpha_t\delta_t\leq\beta_t$ almost surely.

\subsection{Proof of Lemma~\ref{lem:stop-potential}: The potential inequality}\label{subsec:stop-potential-proof}

To establish the desired inequality in rearranged form, we observe that  
\begin{align}
 \exsub{\mathbb P}{\Phi_{t+1}\middle|\Hist_t}
 ~~&=~~ \frac{e}{e-1} \cdot \exsub{\mathbb P}{ (1-\hat q_{t+1})\alpha_{t+1}\middle|\Hist_t} \nonumber \\
 &=~~  \frac{e}{e-1} \cdot (1-\hat q_{t+1}) \cdot  \exsub{\mathbb P}{\alpha_{t+1}\middle|\Hist_t} \label{eqn:stop-potential-proof_1}\\
 &\leq~~  \frac{e}{e-1} \cdot (1-\hat q_{t+1})\alpha_t \label{eqn:stop-potential-proof_2}\\
 &=~~  \Phi_t-\frac{e}{e-1} \cdot \alpha_t\delta_t \label{eqn:stop-potential-proof_3}\\
 &\leq~~  \Phi_t-G_t. \label{eqn:stop-potential-proof_4}
\end{align}
In this derivation, equality~\eqref{eqn:stop-potential-proof_1} is obtained by recalling that $\hat q_{t+1} = \min \{ q_{t+1}, 1 \} = \min \{ z_t, 1 \}$ is already known at stage $t$, and we may therefore take $1-\hat q_{t+1}$ outside the conditional expectation. Inequality~\eqref{eqn:stop-potential-proof_2} holds due to the supermartingale property of $(\alpha_t)$, noting that $1-\hat q_{t+1} \geq 0$. Equality~\eqref{eqn:stop-potential-proof_3} substitutes the definitions of $\Phi_t  = \frac{e}{e-1} \cdot (1-\hat q_t)\alpha_t$ and $\delta_t=\hat q_{t+1}-\hat q_t$. Finally, inequality~\eqref{eqn:stop-potential-proof_4} follows from Lemma~\ref{lem:stop-stage-charge}, stating in particular that $G_t \leq \frac{e}{e-1} \cdot \alpha_t \delta_t$.

\subsection{Proof of Lemma~\ref{lem:stop-tail}: Bounding the remaining cost}\label{subsec:stop-tail-proof}

Let us consider any stopping time $S$ with respect to $(\Hist_t)$, taking values in $\{1,\ldots,n+1\}$. By Lemma~\ref{lem:stop-potential}, we know that $G_t \leq \Phi_t - \exsub{\mathbb P}{\Phi_{t+1}\middle|\Hist_t}$ almost surely, for every $t\in[n]$. Therefore, when multiplying this relation by $\ind\{S\leq t\}$ and taking expectations, we have
\begin{align}
\exsub{\mathbb P}{\ind\{S\leq t\} \cdot G_t}
~~&\leq~~ \exsub{\mathbb P}{ \ind\{S\leq t\} \cdot 
\left( \Phi_t-\exsub{\mathbb P} {\Phi_{t+1}\middle|\Hist_t} \right)} \nonumber \\
&=~~ \exsub{\mathbb P}{ 
\exsub{\mathbb P} { \ind\{S\leq t\} \cdot (\Phi_t- \Phi_{t+1})\middle|\Hist_t} } \nonumber \\
&=~~ \exsub{\mathbb P}{
\ind\{S\leq t\} \cdot (\Phi_t-\Phi_{t+1})}. \label{eqn:stop-tail-proof_1}
\end{align}
Here, the first equality follows from the $\Hist_t$-measurability
of both $\ind\{S\leq t\}$ and $\Phi_t$, since $S$ is a stopping time and since the potential process $(\Phi_t)$ is adapted. 

Now, given this connection, we obtain the desired bound by noting that
\begin{align}
 \exsub{\mathbb P}{\sum_{t=S}^n G_t}
 ~~&=~~
 \sum_{t \in [n]}
 \exsub{\mathbb P}{\ind\{S\leq t\} \cdot G_t} \nonumber \\
 &\leq~~
 \exsub{\mathbb P}{
 \sum_{t \in [n]}
 \ind\{S\leq t\} \cdot (\Phi_t-\Phi_{t+1})} \label{eqn:stop-tail-proof_2} \\
 &=~~
 \exsub{\mathbb P}{\Phi_S-\Phi_{n+1}} \nonumber \\
 &=~~
 \exsub{\mathbb P}{\Phi_S}, \label{eqn:stop-tail-proof_3}
\end{align}
where inequality~\eqref{eqn:stop-tail-proof_2} is obtained by summing~\eqref{eqn:stop-tail-proof_1} over $t\in[n]$, and equality~\eqref{eqn:stop-tail-proof_3} holds since $\Phi_{n+1} = \frac{e}{e-1} \cdot (1-\hat q_{n+1})\alpha_{n+1} =  0$, as the terminal stopping price is $\alpha_{n+1}=0$. For the specific case of $S=n+1$, the sum $\sum_{t=S}^n G_t$ is empty whereas $\Phi_S = \Phi_{n+1} =0$, so the
same calculation remains valid.
\section{The Common-Threshold Inventory Policy}\label{sec:algorithm}

In what follows, we combine the unitwise cost decomposition of
Section~\ref{sec:accounting} with the weighted supermartingale
stopping theorem to construct and analyze our common-threshold
 policy.
The key coordination challenge is to ensure that the stopping decisions for
individual units are compatible, in the sense that a higher supply
label cannot be ordered before a lower one.
We show that  monotonicity properties of the marginal charges allow
us to meet this requirement by sharing one random threshold
across all units. To this end,
Section~\ref{subsec:construction} describes our overall policy, explains
how its clocks incorporate successive observations, and proves
the necessary monotonicity properties.
Section~\ref{subsec:performance} establishes our $\frac{e}{e-1}$-approximation
guarantee by aggregating single-unit comparisons and bounds
the number of conditional-probability oracle queries and arithmetic
operations involved in an efficient implementation.
Finally, Section~\ref{subsec:tightness} presents a family of
single-period instances showing that the  factor $\frac{e}{e-1}$
is tight for this policy.

\subsection{Policy description} \label{subsec:construction}

Our common-threshold policy $\policyct$ first draws a shared threshold, which is held fixed throughout the planning horizon. By contrast, at every time period, new increments to the accumulated clock of each inventory label are computed based on the current observations, while previous ones are kept unchanged. At each ordering opportunity, we identify the labels whose clocks have reached the drawn
threshold and order those that have not already been acquired. These ideas are summarized below, followed by further details on clock evaluation and the feasibility of our policy. 

\begin{tcolorbox}[
    breakable,
    colback=white,
    colframe=black!65,
    boxrule=0.6pt,
    arc=0pt,
    left=9pt,
    right=9pt,
    top=8pt,
    bottom=8pt,
    before skip=12pt,
    after skip=12pt
]
\textbf{The common-threshold policy {\boldmath $\policyct$}}

\begin{enumerate}
    \item \textbf{Initialization.}
    Draw the random threshold $\rho$ as explained
    in Section~\ref{subsec:stoprule}, independently of demands and
    signals, and initialize $Q_{[0]}=0$.

    \item \textbf{Ordering decisions.}
    At the beginning of each period $t=1,\ldots,n$, perform the
    following steps prior to observing the current demand $D_t$:
    \begin{enumerate}
        \item \emph{Update  current history.}
        Observe the past demands and newly available signals,
        summarized by $\Hist_t$.

        \item \emph{Evaluate candidate labels.}
        For each  candidate label $k$ tested by the binary
        search in item~(c), compute $z_t(k)= \sum_{s \leq t} \frac{\beta_s(k)}{\alpha_s(k)}$
        by evaluating $\alpha_s(k)$ and $\beta_s(k)$ conditional
        on their original observed histories $\Hist_s$.
        Declare $k$ passing if $z_t(k)\geq\rho$.

        \item \emph{Place new order.}
        Find the largest passing label by binary search over
        $\{Q_{[t-1]},\ldots,M\}$.
        Set $Q_{[t]}$ to this label,
        and place an order of size $Q_t = Q_{[t]}-Q_{[t-1]}$.
    \end{enumerate}

    \item \textbf{Remaining periods.}
    Place no new orders after period $n$, i.e., set $Q_t=0$ for $t=n+1,\ldots,T$.
\end{enumerate}
\end{tcolorbox}

\paragraph{Clock updates.} To elaborate on how candidate labels are evaluated in item~(b), let us recall that for every label $u$, its clock through period $t$ is given by $z_t(u) = \sum_{s \leq t} \frac{\beta_s(u)}{\alpha_s(u)}$. According to definitions~\eqref{eq:a} and~\eqref{eq:b}, these charges are specified by
\begin{equation} \label{eqn:algorithm_alpha_beta}
\alpha_s(u) ~~=~~ \exsub{\omega}{  \sum_{r=s+L}^T h_r \cdot \ind\{D_{[r]}<u\} \middle| \Hist_s} \quad \text{and} \quad 
 \beta_s(u) ~~=~~ p_{s+L} \cdot \prsub{\omega}{D_{[s+L]}\geq u \middle| \Hist_s},
\end{equation}
meaning that they are conditional on the history $\Hist_s$ available at the beginning of period $s \leq t$. Thus, a forecast in period $t$ contributes the new ratio $\frac{ \beta_t(u) }{ \alpha_t(u) }$, without affecting previously accumulated increments. For example, the clock in period $2$ is $\frac{ \beta_1(u) }{ \alpha_1(u) }+ \frac{ \beta_2(u) }{ \alpha_2(u) }$, where the first ratio $\frac{ \beta_1(u) }{ \alpha_1(u) }$ is based on the period-$1$ forecast, while the second ratio $\frac{ \beta_2(u) }{ \alpha_2(u) }$ is based on that of period $2$. It is important to emphasize that recomputing the first ratio using period-$2$ information would
define a different clock process, for which the approximation guarantee of
Theorem~\ref{thm:stopping} would not follow from our analysis.

\paragraph{Ordering feasibility.} Yet another important comment is concerned with two monotonicity properties
required to justify item~(c), which are established in Lemma~\ref{lem:prefix} below. First, for each fixed period $t$, we prove that the accumulated clock $z_t(u)$ is nonincreasing in $u$, implying that its new passing
labels form a prefix of $Q_{[t-1]},\ldots,M$, and thus, the largest such label can be determined by binary search.
Second, for each fixed label $u$, we prove that its clock $z_t(u)$ is nondecreasing in $t$, meaning that every
previously passing label continues to pass.
Consequently, the purchased prefixes $1,\ldots,Q_{[t]}$ can only grow, i.e., $Q_{[t]}\geq Q_{[t-1]}$, and every order quantity $Q_t=Q_{[t]}-Q_{[t-1]}$ is indeed nonnegative.

\begin{lemma} \label{lem:prefix}
At every observed history, $z_t(u)$ is nonincreasing in $u$. Along a realized history, $z_t(u)$ is nondecreasing in $t$. 
\end{lemma}
\begin{proof}
Let us fix a period $s$ and two labels $u\leq u'$. As explained at the end of Section~\ref{subsec:supermart}, the stopping prices and waiting charges of these labels satisfy $\alpha_s(u)\leq \alpha_s(u')$ and $\beta_s(u)\geq \beta_s(u')$, respectively. As such, we show that $\frac{\beta_s(u)}{\alpha_s(u)} \geq\frac{\beta_s(u')}{\alpha_s(u')}$ by considering three cases:
\begin{itemize}
    \item {\em When $\alpha_s(u) > 0$}: Here, $\frac{\beta_s(u)}{\alpha_s(u)}  \geq\frac{\beta_s(u')}{\alpha_s(u)}  \geq\frac{\beta_s(u')}{\alpha_s(u')}$, given that $\alpha_s(u)\leq \alpha_s(u')$ and $\beta_s(u)\geq \beta_s(u')$.

    \item {\em When $\alpha_s(u) = 0$ and $\beta_s(u) > 0$}: In this case, by recalling the conventions $\frac{ 0 }{ 0 }=0$ and $\frac{ \beta }{ 0 } =\infty$ for $\beta>0$, it follows that $\frac{\beta_s(u)}{\alpha_s(u)} = \infty \geq\frac{\beta_s(u')}{\alpha_s(u')}$.

    \item {\em When $\alpha_s(u) = 0$ and $\beta_s(u) = 0$}: Since $\beta_s(u') \leq \beta_s(u) = 0$, we have $\frac{\beta_s(u')}{\alpha_s(u')} = 0 \leq \frac{\beta_s(u)}{\alpha_s(u)}$.
\end{itemize}
By summing the relation $\frac{\beta_s(u)}{\alpha_s(u)} \geq\frac{\beta_s(u')}{\alpha_s(u')}$ over $s\leq t$, we conclude that $z_t(u) \geq z_t(u')$, namely, $z_t(u)$ is nonincreasing in $u$. To obtain the second property, stating that $z_t(u)$ is nondecreasing in $t$, we simply note that $z_{t+1}(u) = z_t(u) + \frac{\beta_{t+1}(u)}{\alpha_{t+1}(u)} \geq z_t(u)$, since the new increment $\frac{\beta_{t+1}(u)}{\alpha_{t+1}(u)}$ is nonnegative.
\end{proof}

\subsection{Performance analysis}\label{subsec:performance}

To finalize our main result, as formally stated in Theorem~\ref{thm:main}, we argue that against every nonanticipatory policy $P$, our common-threshold policy $\policyct$ guarantees an expected total cost of
\begin{equation} \label{eqn:performance_comp}
\exsub{\omega,\rho}{F(\policyct)} ~~\leq~~ \frac{e}{e-1} \cdot \exsub{\omega,\xi_P}{F(P)}.
\end{equation}
Moreover, we will explain how this policy can be implemented through $O(T^3\log M)$ conditional-probability oracle queries and arithmetic operations overall.

\paragraph{Approximation guarantee.} Given the unitwise cost decomposition prescribed by Lemma~\ref{lem:unitcost}, the expected cost of any replenishment policy $P$ can be written as
\[  \exsub{\omega,\xi_P}{F(P)} ~~=~~ K_0+\sum_{u \in [M]}\exsub{\omega,\xi_P}{\alpha_{S_P(u)}(u)        +\sum_{t<S_P(u)}\beta_t(u)}. \]
The exact same decomposition applies to our common-threshold policy. Consequently, to derive the desired upper bound~\eqref{eqn:performance_comp}, it suffices to prove a single-unit inequality, showing that for every inventory label $u$, 
\begin{equation} \label{eqn:global_policy_u}
\exsub{\omega,\rho}{\alpha_{S_{\policyct^\rho}(u)}(u)+\sum_{t<S_{\policyct^\rho}(u)}\beta_t(u)}
 ~~\leq~~ \frac{e}{e-1} \cdot \exsub{\omega,\xi_P}{\alpha_{S_P(u)}(u)+\sum_{t<S_P(u)}\beta_t(u)}.    
\end{equation}
Here, $\policyct^\rho$ denotes our common-threshold policy
with its threshold fixed at $\rho$, and $S_{\policyct^\rho}(u)$
designates the period in which this policy orders unit $u$, with
value $n+1$ when this unit is never ordered. In other words, by definition~\eqref{eqn:stopping_policy_unit},
\[
 S_{\policyct^\rho}(u)
 ~~=~~
 \min(
 \{t\in[n]:Q_{[t]}\geq u\}\cup\{n+1\}
 )
 ~~=~~
 \min(
 \{t\in[n]:z_t(u)\geq\rho\}\cup\{n+1\}
 ),
\]
where the second equality follows from Lemma~\ref{lem:prefix}. Specifically,
at each period $t$, the labels whose clocks have reached
$\rho$ form the prefix $1,\ldots,Q_{[t]}$,
implying that $Q_{[t]}\geq u$ if and only if $z_t(u)\geq\rho$. Crucially, the weighted stopping rule of Section~\ref{subsec:stoprule},
applied to the processes $\alpha_t(u)$ and $\beta_t(u)$
with the same threshold $\rho$, stops at the first period
satisfying this very condition.
Therefore, its stopping time $\tau_u(\rho)$ coincides
with $S_{\policyct^\rho}(u)$, including the value $n+1$
when no crossing occurs.

To apply Theorem~\ref{thm:stopping}, we recall that
Lemma~\ref{lem:supermart} establishes the required nonnegative
supermartingale property of the stopping prices $\alpha_t(u)$,
while the waiting charges $\beta_t(u)$ are nonnegative,
adapted, and integrable by their definition.
If the comparison policy $P$ is randomized, we apply Theorem~\ref{thm:stopping}
with the filtration enlarged by its independent seed $\xi_P$,
choosing $\rho$ independently of both this seed and the
exogenous process.
This enlargement makes $S_P(u)$ a stopping time and,
by independence, leaves the conditional charges and their
supermartingale property unchanged. This conclusion is precisely what we are seeing in inequality~\eqref{eqn:global_policy_u}, with expanded forms of $\mathcal{C}(\tau_u(\rho))$ on the left and $\mathcal{C}(S_P(u))$ on the right.

\paragraph{Conditional-probability implementation.} Circling back to Section~\ref{subsec:model}, we remind readers that, to deal with arbitrarily structured demand and signal processes, we assume exact oracle access to cumulative-demand probabilities conditional on observed histories. In particular, at the beginning of any period $t$, our oracle  receives as input a history index $s\in\{1,\ldots,t\}$, a target period $r\in\{s,\ldots,T\}$, and  threshold $u\in[M]$, returning the realized value of $H_{s,r}(u) = \prsubpar{\omega}{D_{[r]}<u|\Hist_s}$. This expression stands for the probability that cumulative demand through period $r$ is strictly below $u$, given the information available at the beginning of period $s$. Given this oracle access, we observe that for every label $u$, representation~\eqref{eqn:algorithm_alpha_beta} of the stopping price $\alpha_s(u)$ and waiting charge $\beta_s(u)$ allows us to express these terms as
\begin{equation} \label{eq:oraclecharges}
\alpha_s(u) ~~=~~ \sum_{r=s+L}^T h_r \cdot H_{s,r}(u) \qquad \text{and} \qquad 
 \beta_s(u) ~~=~~ p_{s+L} \cdot ( 1-H_{s,s+L}(u) ).
\end{equation}

We proceed by arguing that our policy can be implemented through $O(T^3\log M)$ conditional-probability oracle queries and arithmetic operations overall. To this end, let us fix a period $t$ and a candidate label $k$.
For each $s\leq t$, evaluating $\alpha_s(k)$ and $\beta_s(k)$
through equation~\eqref{eq:oraclecharges} requires $O(T)$ oracle
queries and $O(T)$ arithmetic operations.
Consequently, computing the accumulated clock $z_t(k) = \sum_{s \leq t} \frac{ \beta_s(k) }{ \alpha_s(k) }$
requires $O(tT)$ such operations. By Lemma~\ref{lem:prefix}, the passing labels form a prefix, and every previously passing label continues to pass. Therefore, binary search over $\{Q_{[t-1]},\ldots,M\}$ determines the new
cumulative order quantity $Q_{[t]}$ through $O(\log M )$ candidate evaluations, and thus corresponds to $O(tT\log M)$ oracle
queries and arithmetic operations within this time period. Summing over all ordering periods $t \in [n]$ leads to an overall running time of $O(T^3\log M)$ under our conditional-probability oracle assumption.

\subsection{Tight example}\label{subsec:tightness}

In what follows, we describe a parametric family $(I_{\eps})_{\eps \in (0,1/2)}$ of single-period instances, for which
the ratio between our policy's expected cost and the optimal
expected cost tends to $\frac{e}{e-1}$ as $\eps$ tends to zero.

\paragraph{Instance construction.} Given a parameter $\eps\in(0, \frac{ 1 }{ 2 } )$, we consider a  single-period instance with zero lead time, i.e., $T = 1$ and $L = 0$. Consequently, any order placed at the beginning of period $1$ arrives
immediately, before demand is observed. The per-unit holding and backlog costs are set to $h_1=p_1=1$. The demand $D_1$ in this period is a Bernoulli random variable, taking the values $1$ and $0$ with respective probabilities $\eps$ and $1 - \eps$. Thus, we may fix $M=1$ and restrict the order quantity to
$Q_1\in\{0,1\}$. Finally, there are no informative signals before our ordering decision, meaning that any policy knows the parameter $\eps$ and the resulting demand distribution, but it does not know which of the two demand realizations will
occur. 

\paragraph{The optimal policy.} If no order is placed, we incur a backlog cost of $1$ when $D_1=1$ and no cost otherwise; hence, this decision leads to an expected cost of
$\eps$. If one unit is ordered, a holding
cost of $1$ is incurred when $D_1=0$, and our expected cost is $1-\eps$. Since $\eps< \frac{ 1 }{ 2 }$, placing no order
is strictly preferable and $\opt = \eps$. In addition, randomization cannot improve this choice. Indeed, consider any
feasible policy $P$ that orders a single unit with probability
$\theta\in[0,1]$ and orders nothing otherwise. Since its ordering
decision is independent of the demand realization, 
\[
 \exsub{\omega,\xi_P}{F(P)}
 ~~=~~
 (1-\theta) \cdot \eps+\theta \cdot (1-\eps)
 ~~=~~
 \eps+\theta \cdot (1-2\eps)
 ~~\geq~~
 \eps.
\]

\paragraph{The common-threshold policy.}
We now examine how our policy behaves on this instance. For the only relevant inventory label $u=1$, its stopping price and
waiting charge are
\[
 \alpha_1(1)
 ~~=~~
 h_1\cdot\prsub{\omega}{D_1<1}
 ~~=~~
 1-\eps
 \qquad \text{and} \qquad
 \beta_1(1)
 ~~=~~
 p_1\cdot\prsub{\omega}{D_1\geq1}
 ~~=~~
 \eps.
\]
Consequently, the accumulated clock at the only ordering opportunity is $z_1(1) = 
 \frac{\beta_1(1)}{\alpha_1(1)} = \frac{\eps}{1-\eps}$. Our policy orders one unit precisely when the random threshold
$\rho$ is at most this clock value; otherwise, it places no
order. Since $\eps\in(0,\frac{1}{2})$, we have $z_1(1)\in(0,1)$, and  the probability of ordering one unit is $F_{\rho}( z_1(1) ) =  \frac{e^{\eps/(1-\eps)}-1}{e-1}$. Since our random threshold $\rho$ is independent of demand, the preceding
cost calculation implies that 
\[ \exsub{\omega,\rho}{F(\policyct)} ~~=~~ 
 (1-F_{\rho}(z_1(1))) \cdot \eps
 +F_{\rho}(z_1(1))  \cdot (1-\eps). \]

\paragraph{Bounding the expected-cost
ratio.} On the one hand,
\begin{align*}
 \frac{\exsub{\omega,\rho}{F(\policyct)}}{\opt}
 ~~&=~~
 \frac{
 (1-F_{\rho}(z_1(1))) \cdot \eps
 +F_{\rho}(z_1(1))  \cdot (1-\eps)}{\eps}\\
 &=~~
 1+\frac{1-2\eps}{e-1}
 \cdot\frac{e^{\eps/(1-\eps)}-1}{\eps}\\
 &\leq~~
 1+\frac{1-2\eps}{e-1}
 \cdot\frac{1}{\eps}
 \cdot\frac{\eps}{1-2\eps}\\
 &=~~
 \frac{e}{e-1},
\end{align*}
where the sole inequality above holds since $e^x-1\leq\frac{x}{1-x}$ for $0\leq x<1$. For the reverse bound, we have
\begin{align*}
 \frac{\exsub{\omega,\rho}{F(\policyct)}}{\opt}
 ~~&=~~
 1+\frac{1-2\eps}{e-1}
 \cdot\frac{e^{\eps/(1-\eps)}-1}{\eps}\\
 &\geq~~
 1+\frac{1-2\eps}{e-1}
 \cdot\frac{1}{1-\eps}\\
 &=~~
 \frac{e}{e-1}
 -\frac{\eps}{(e-1)(1-\eps)}\\
 &\geq~~
 \frac{e}{e-1}-\frac{2\eps}{e-1},
\end{align*}
where the first inequality follows from $e^x-1\geq x$. Combining the two bounds, we obtain
\[
 \frac{e}{e-1}-\frac{2\eps}{e-1}
 ~~\leq~~
 \frac{\exsub{\omega,\rho}{F(\policyct)}}{\opt}
 ~~\leq~~
 \frac{e}{e-1}.
\]
Letting $\eps$ tend to zero shows that the expected-cost
ratio converges to $\frac{e}{e-1}$. Thus, the approximation
guarantee of our common-threshold policy is tight, even
for single-period instances.
\section{Concluding Remarks}\label{sec:conclusions}

We conclude this work by highlighting a number of open questions and directions for future research. Beyond the natural question of improved guarantees, we discuss how our stopping formulation might yield stronger approximability results when replenishment decisions interact through setup costs, capacity constraints, or multiple echelons.

\paragraph{The optimal approximation guarantee.} While still addressing general demand correlations and evolving forecasts, can we surpass our $\frac{ e }{ e - 1 }$-approximation for periodic-review stochastic inventory control? The single-period Bernoulli example in Section~\ref{subsec:tightness} establishes
tightness for our common-threshold policy, leaving open the performance attainable by alternative polynomial-time policies. Toward this goal, a promising avenue could be to exploit specific structural properties of
the weighted stopping problems generated by marginal-cost accounting, as their prices and waiting charges are not arbitrary, but rather arise from the same cumulative
demand process and satisfy compatibility conditions across labels. Can these relationships support a stronger comparison than separately treating each unit? For example, we might combine threshold decisions with direct optimization of selected replenishment decisions. Conversely, a matching lower bound under the computational model of Section~\ref{subsec:model} would identify the fundamental approximation limit for this general demand setting.

\paragraph{Setup costs and coordinated replenishment.} Can we obtain an approximation factor strictly below $3$ for
stochastic lot sizing with positive lead times, improving on the work of \citet{lot2013} under the same demand assumptions? In essence, incurring a fixed cost whenever an order is placed creates an incentive to replenish multiple units together. Our stopping formulation suggests studying blocks of consecutive labels, with each block sharing a single setup cost while its units retain their individual holding and backlogging charges. The central question is how to select these blocks and their ordering times as forecasts evolve. In particular, placing an order for urgent units may justify including other units whose individual clocks have not yet crossed their thresholds. Can we jointly account  for this advance replenishment and the setup costs saved by combining orders? Such an analysis should charge each setup cost only once while preserving enough of the unitwise stopping guarantee to obtain a factor below $3$.

\paragraph{Capacity constraints and competing units.} Can we improve the approximation guarantee of $2$, attained by \citet{capacitated2008} for capacitated stochastic inventory
control, even when the ordering capacity is constant across periods?
Our unitwise formulation suggests viewing capacity as a resource
allocated among competing stopping decisions.
Several labels may simultaneously become eligible for acquisition (i.e., have threshold-crossing clocks),
while the enforced capacity permits ordering only some of them.
Moreover, currently unused capacity  cannot be carried forward, meaning that allocation rules must account for future
shortages as well as current waiting charges.
One promising approach could be to supplement unit clocks with adapted shadow prices for scarce ordering capacity.
Can we construct these prices and a joint potential that compare
the resulting policy with the optimal capacitated policy?
A further objective is to extend such a comparison to multiple
items sharing a common per-period ordering capacity, allowing correlated demands
across both items and periods.

\paragraph{Coordinating decisions across multiple echelons.} Can our approach improve the approximation factors between \(2\) and \(3\) established by \citet{multiechelon2017}
for serial inventory systems under different assumptions?
In such a system, each unit undergoes a sequence of replenishment
decisions as it moves toward customers.
These decisions must respect transportation lead times and the
availability of supply at upstream locations.
We therefore ask whether the ordering time assigned to each unit
can be replaced by a sequence of compatible stopping times,
with charges that account for its progress through the system. Here, compatibility ensures that each transfer decision is based on currently available information, and a unit cannot leave an intermediate location before arriving there. Perhaps a joint potential could compare these decisions across all
echelons without incurring repeated approximation losses.
The main challenge is to coordinate the benefit of reducing customer backlogs or making stock available sooner with the holding costs incurred along its route.

\phantomsection
\addcontentsline{toc}{section}{References}
\bibliographystyle{plainnat}
\bibliography{BIB-Inventory}

\addtocontents{toc}{\protect\setcounter{tocdepth}{2}} 
\appendix
\section{Extension 1: Unbounded Demand}
\label{app:unbounded_demand}

In what follows, we extend our common-threshold policy and its approximation guarantee to unbounded integer-valued demand. Beyond previously encountered obstacles, we must ensure that only
finitely many units are ordered, in spite of facing an infinite sequence of potential labels. To this end, Appendix~\ref{subsec:unbounded-setting} formalizes the current demand model and states our  performance guarantee.
Appendix~\ref{subsec:unbounded-policy} explains how the common-threshold policy can be adapted to this setting, which requires replacing demand-bounded search by a doubling procedure. Finally, Appendix~\ref{subsec:unbounded-analysis} establishes feasibility and derives our approximation guarantee.

\subsection{Model specifics and main result}
\label{subsec:unbounded-setting}

\paragraph{Demand assumptions.}
We retain the periodic-review model of Section~\ref{subsec:model},
including integer demands and order quantities, a deterministic lead
time, and exogenous demands and signals that may be arbitrarily
correlated and nonstationary.
The initial inventory,  backlog, and  pipeline remain
empty, and there are no purchase costs; these features can be incorporated via standard ideas, as explained in Section~\ref{subsec:contributions}. 
However, we no longer assume that the total demand $D_{[T]}$ is almost surely bounded by a known integer $M$. Instead, similarly to \citet{lprs2007}, we operate under the assumption that, while its support could be unbounded, total demand still has finite expectation (i.e., $\exsubpar{\omega}{D_{[T]}}<\infty$) and that inventory left when the planning horizon ends incurs a
strictly positive holding cost, $h_T>0$. 

In this setting, a feasible policy places nonnegative integer orders that are finite almost surely, depending only on the observed history and an independent random seed. In terms of comparisons, we are still competing against the fully adaptive benchmark, $\opt= \inf_{P\in\PP}\exsub{\omega,\xi_P}{F(P)}$, where $\PP$ stands for the family of computationally unbounded nonanticipatory policies  for the periodic-review model under consideration.

\paragraph{Main result.} We construct a modified common-threshold policy, $\policyunb$, placing finite orders almost surely, with an expected total cost of $\exsubpar{\omega,\rho}{F(\policyunb)}  \leq \frac{e}{e-1} \cdot \opt$. This policy can be implemented by finitely many conditional-probability queries almost surely, noting that the number of such queries depends on the realized cumulative order quantities.

\subsection{Policy description}
\label{subsec:unbounded-policy}

Similarly to Section~\ref{subsec:construction}, our modified common-threshold policy $\policyunb$ draws a shared threshold
$\rho$. At each ordering opportunity, we identify the labels whose clocks have reached the drawn
threshold and order those that have not already been acquired. Nevertheless, without a known upper bound $M$, the main algorithmic difference resides in locating the next label that fails the crossing test. As detailed below, this task is performed by repeatedly doubling a candidate endpoint and then conducting a binary search over the resulting finite interval.

\begin{tcolorbox}[
    breakable,
    colback=white,
    colframe=black!65,
    boxrule=0.6pt,
    arc=0pt,
    left=9pt,
    right=9pt,
    top=8pt,
    bottom=8pt,
    before skip=12pt,
    after skip=12pt
]
\textbf{The modified common-threshold policy {\boldmath $\policyunb$}}

\medskip Follow the common-threshold policy of Section~\ref{subsec:construction}, replacing its bounded search by a doubling procedure at each ordering
period $t$. Specifically, to determine the largest passing label across $Q_{[t-1]}, Q_{[t-1]}+1, \ldots$:
\begin{enumerate}
    \item \textbf{Locate a failing upper endpoint.}
    Set $a=Q_{[t-1]}$ and $b=\max\{1,2a\}$.
    While $b$ passes, update $(a,b)\leftarrow(b,2b)$.
    Upon termination, $a$ passes and $b$ fails.

    \item \textbf{Locate the ordering boundary.}
    Apply binary search between $a$ and $b$ to determine the largest
    passing index, and set $Q_{[t]}$ to this index.
\end{enumerate}
\end{tcolorbox}

\paragraph{Further interpretation.} It is worth pointing out that the pairwise comparisons within the proof of Lemma~\ref{lem:prefix} do not depend on assuming bounded demand. Thus, for each ordering period $t$, the accumulated clock $z_t(u)$ is nonincreasing in $u$. Moreover, each label's clock is nondecreasing over time, since its increments are nonnegative. The passing labels therefore form a growing prefix, and whenever this prefix is finite, our search identifies its largest label, corresponding to a cumulative order quantity of 
\begin{equation}\label{eq:unbounded-orders}
Q_{[t]} ~~=~~  \max \left(  \{u\in\{1 ,2 , \ldots \} : z_t(u) \geq \rho \} \cup \{0\} \right).
\end{equation}
The next section will argue that a failing upper endpoint exists almost surely,
which justifies both the doubling procedure above and representation~\eqref{eq:unbounded-orders} of our cumulative order quantity.

\subsection{Analysis}
\label{subsec:unbounded-analysis}

\paragraph{Finite passing prefixes.}
We first prove that the passing prefix is finite at every ordering
period $t$ almost surely. To establish this claim, we show that $z_t(u)\to0$ almost surely as the label $u$ tends to infinity. Whenever this convergence holds, any positive threshold $\rho$ admits a finite cutoff $U_t$ such that $z_t(u)<\rho$ for every $u\geq U_t$. Hence, every passing label belongs to the set $\{1,\ldots,U_t-1\}$. Our doubling procedure must therefore encounter a failing endpoint, after which binary search identifies the ordering boundary. Since $\rho>0$ almost surely, this observation establishes almost-sure termination and finite order quantities at every ordering period.

To argue about the limiting behavior of $z_t(u)$, note that since $\exsubpar{\omega}{D_{[T]}}<\infty$, total demand is finite almost surely.
Consequently, for every period $r$,
the events $\{D_{[r]}<u\}$ increase to an event of probability $1$
as $u$ tends to infinity.
Conditioning on the history $\Hist_s$ at period $s \leq r$ and applying conditional monotone convergence, we have
\[
 H_{s,r}(u)
 ~~=~~
 \prsub{\omega}{D_{[r]}<u\middle|\Hist_s} ~~\convas~~ 1.
\]
In other words, conditional on the information available at period
$s$, cumulative demand through period $r$ falls below a sufficiently
large label with probability approaching $1$. Since there are only finitely many pairs $(s,r)$ with $1\leq s\leq r\leq T$, these limits hold simultaneously outside a single event of probability zero.
We work on the remaining realizations throughout the following
argument.

By definition~\eqref{eq:a} of the stopping prices, for every period $s\in[n]$,
\[ \alpha_s(u) ~~=~~ \exsub{\omega}{  \sum_{r=s+L}^T h_r \cdot \ind\{D_{[r]}<u\} \middle| \Hist_s} ~~=~~ \sum_{r=s+L}^T h_r \cdot H_{s,r}(u) ~~\longrightarrow~~ \sum_{r=s+L}^T h_r ~~\geq~~ h_T ~~>~~ 0.
\]
Similarly, by definition~\eqref{eq:b} of the waiting charges, 
\[ \beta_s(u) ~~=~~ p_{s+L} \cdot \prsub{\omega}{D_{[s+L]}\geq u \middle| \Hist_s} 
~~=~~ p_{s+L} \cdot (1-H_{s,s+L}(u))
 ~~\longrightarrow~~ 0. \]
Thus, for sufficiently large labels, the expected holding charge $\alpha_s(u)$ remains above $\frac{ h_T }{ 2 }$, while the expected backlog charge $\beta_s(u)$ vanishes. Therefore,
\[ 0 ~~\leq~~ \frac{\beta_s(u)}{\alpha_s(u)} ~~\leq~~ 
 \frac{2}{h_T} \cdot \beta_s(u) ~~\longrightarrow~~ 0. \]
Since the accumulated clock $z_t(u)
 = \sum_{s \leq t}\frac{\beta_s(u)}{\alpha_s(u)}$ contains only finitely many such
increments, we obtain $z_t(u)\to0$ almost surely, as desired.

\paragraph{Feasibility and unit ordering times.} Given that passing prefixes can only grow, the cumulative order quantities
in representation~\eqref{eq:unbounded-orders} are nondecreasing.
The resulting orders are therefore nonnegative integers, which are finite almost surely by the preceding argument. Moreover, each candidate evaluation utilizes only previously observed
histories and the independent threshold $\rho$, confirming that $\policyunb$ is a nonanticipatory policy. As in Section~\ref{subsec:performance}, the ordering time $S_{ \policyunb^\rho }( u )$ of each label $u$ agrees with its weighted stopping time $\tau_u(\rho)$, namely,
\[
 S_{\policyunb^\rho}(u)
 ~~=~~
 \min\left(
       \{t\in[n]:z_t(u)\geq\rho\}\cup\{n+1\}
     \right)
 ~~=~~
 \tau_u(\rho),
\]
since $Q_{[t]} \geq u$ holds exactly when $z_t(u)\geq\rho$. This equivalence includes labels that never cross the threshold,
whose ordering time is $n+1$.

\paragraph{Extended cost decomposition.} Our next objective is to bound the expected total cost of $\policyunb$ in relation to $\opt$ by following the unitwise comparison of Section~\ref{subsec:performance}. For this purpose, the important observation is that the cost decomposition of Lemma~\ref{lem:unitcost} extends 
to the countably infinite label set.
Its conditioning argument for each fixed label remains unchanged, and nonnegativity allows us to interchange the infinite sum and
expectation by Tonelli's theorem.
Thus, for every feasible policy $P$,
\begin{equation} \label{eq:unbounded-unitcost}
 \exsub{\omega,\xi_P}{F(P)} ~~=~~ K_0+\sum_{u =1}^{ \infty }\exsub{\omega,\xi_P}{\alpha_{S_P(u)}(u)        +\sum_{t<S_P(u)}\beta_t(u)},
\end{equation}
noting that this identity remains valid when either side is infinite. Here, the policy-independent initial backlog term $K_0=\exsubpar{\omega}{\sum_{t \in [L]} p_tD_{[t]}}$ remains finite, again by the assumption that $\exsubpar{\omega}{D_{[T]}}<\infty$. 

\paragraph{Approximation guarantee.} To exploit this cost accounting scheme, we observe that the supermartingale property established in Lemma~\ref{lem:supermart} applies regardless of whether demand is bounded. Thus, Theorem~\ref{thm:stopping} leads to the same unitwise comparison as in Section~\ref{subsec:performance}, including against randomized
policies through independent-seed enlargement. Summing these comparisons over all labels, applying decomposition~\eqref{eq:unbounded-unitcost}, and recalling that $S_{\policyunb^\rho}(u) = \tau_u(\rho)$, we conclude that
\[  \exsub{\omega,\rho}{F(\policyunb)}
 ~~ \leq ~~ K_0+\frac{e}{e-1} \cdot 
 \left( \exsub{\omega,\xi_P}{F(P)}-K_0 \right) ~~ \leq ~~ \frac{e}{e-1} \cdot \exsub{\omega,\xi_P}{F(P)} 
 \] 
for every feasible policy $P$ with finite expected cost. Such a policy clearly exists, since never ordering incurs expected cost at most
$(\sum_{t \in [T]} p_t) \cdot \exsubpar{\omega}{D_{[T]}} < \infty$. Finally, taking the infimum over all finite-cost feasible policies yields the promised approximation guarantee, $\exsubpar{\omega,\rho}{F(\policyunb)} \leq \frac{e}{e-1} \cdot \opt$.

\paragraph{Why $\exsubpar{\omega}{D_{[T]}}<\infty$ alone is insufficient.} A close inspection of our analysis reveals that, to keep the stopping prices of large
labels bounded away from zero, we have made explicit use of the assumption that remaining inventory at the end of our planning horizon incurs a strictly positive holding cost. It is natural to ask whether finite expected total demand alone
could suffice for a comparable guarantee, perhaps through a
different policy.
The following example shows that, without an additional condition,
no finite multiplicative approximation is possible
against our benchmark, regardless of computation time.

Let us consider an instance consisting of a single period, with zero lead time, no informative signals, holding cost $h_1 = 0$, and backlog cost $p_1 = 1$. Suppose that the demand $D_1$ in this period is geometrically distributed, with success probability $\frac{ 1 }{ 2 }$, namely $\prpar{D_1=d}=\frac{ 1 }{ 2^d }$ for every positive integer $d$. Thus, demand has unbounded support but finite expectation
$\expar{D_1}=2$.
For any integer $q \geq 0$, the policy $P_q$ that orders
$q$ units incurs expected cost
\[ \exsub{\omega}{F(P_q)} ~~=~~
 \ex{[D_1-q]^+} ~~=~~
 \sum_{d=q+1}^{\infty} \frac{ d-q }{ 2^{d} } ~~=~~ \frac{ 1 }{ 2^{q} } \cdot \sum_{k=1}^{\infty} \frac{ k }{ 2^{k} }~~=~~ \frac{ \expar{D_1} }{ 2^q }  ~~=~~
 \frac{ 1 }{ 2^{q-1} }. \]
This cost is strictly positive for every finite $q$, but converges
to zero as $q$ tends to infinity, implying that $\opt= 0$. Randomization cannot attain this benchmark either. Indeed, with no informative signals, any randomized order quantity
$Q$ is independent of demand. By conditioning on the policy's seed $\xi_P$, we have $\exsubpar{\omega,\xi_P}{F(P)} =  \exsubpar{\xi_P}{\frac{ 1 }{ 2^{Q-1} }}$, which is strictly positive whenever $Q$ is finite almost surely.
Consequently, no such policy can satisfy
$\exsubpar{\omega,\xi_P}{F(P)}\leq c \cdot \opt$ for any finite factor $c$. The created obstruction is that the benchmark $\opt$ is an unattained zero infimum, approached only by ordering quantities that grow without bound.

\section{Extension 2: Divisible Quantities}
\label{app:divisible}

In what follows, we show that our common-threshold policy extends to real-valued demand and divisible orders, primarily by replacing unit sums with integrals over continuous labels. For this purpose, Appendix~\ref{subsec:divisible-setting} specifies the precise model considered and states the performance guarantee we attain.
Appendix~\ref{subsec:divisible-policy} describes how our ordering rule can be adapted by representing cumulative supply as an interval of
labels, whose length measures the quantity acquired. Finally, Appendix~\ref{subsec:divisible-analysis} explains how cost decomposition operates in the continuous setting and establishes the resulting approximation guarantee.

\subsection{Model specifics and main result}
\label{subsec:divisible-setting}

\paragraph{Model, oracle assumption, and benchmark.} Once again, we retain the periodic-review model of Section~\ref{subsec:model},
including its deterministic lead
time, exogenous demands and signals, empty initial state, and linear holding and backlog costs, with no purchase or fixed ordering costs. That said, demands and order quantities are now allowed to take nonnegative real values, still assuming a total demand of $D_{[T]} \leq M$ almost surely, for a known
real bound $M>0$. As such, cumulative orders can be restricted to $[0,M]$, since supply beyond
$M$ cannot satisfy additional demand and can only increase holding
costs.

In this setting, we are competing against the fully adaptive benchmark, $\opt_{\rm div} = \inf_{P\in\PP_{\rm div} }\exsub{\omega,\xi_P}{F(P)}$, where $\PP_{\rm div}$ designates the family of computationally unbounded nonanticipatory policies with cumulative order quantities in \([0,M]\). For oracle access, we extend the basic assumption of Section~\ref{subsec:model} to account for cumulative-demand probabilities with respect to real-valued labels. In other words, at the beginning of period $t$, given a history index $s\in\{1,\ldots,t\}$, a target period $r\in\{s,\ldots,T\}$, and threshold $u\in[0,M]$, our oracle returns the probability $\prsubpar{\omega}{D_{[r]}<u|\Hist_s}$ that cumulative demand through period $r$ is strictly below $u$, given the information available at the beginning of period $s$. 

\paragraph{Main result.} In the remainder of this section, we devise a modified common-threshold policy for divisible quantities, $\policydiv$, attaining an expected total cost of $\exsubpar{\omega,\rho}{F(\policydiv)}  \leq \frac{e}{e-1} \cdot \opt_{\rm div}$. 
As in the integer model, we draw a single threshold $\rho$
independently of demands and signals, which is kept unchanged throughout the planning horizon. The crux of our analysis will reside in observing that, with the appropriate corrections, weighted supermartingale stopping applies to real labels, and integrating these comparisons to recover the expected total cost.

\subsection{Policy description}
\label{subsec:divisible-policy}

We proceed by describing how our common-threshold policy determines ordering quantities when supply is divisible. In the integer model, cumulative orders equal the number of labels whose clocks have reached the shared threshold. Here, rather than counting labels, we measure the length of the interval they occupy. For this purpose, integer labels are replaced by $(0,M]$, where a cumulative order quantity $q$ acquires the interval $(0,q]$. Therefore, increasing cumulative orders from $q$ to $q'$  acquires
$(q,q']$, corresponding to a new order of size $q'-q$.

For each real label $u$, we retain the stopping prices and waiting charges in definitions~\eqref{eq:a} and~\eqref{eq:b}, with its accumulated clock at period $t$ still given by $z_t(u) = \sum_{s \leq t} \frac{\beta_s(u)}{\alpha_s(u)}$. As before, we 
declare a label passing when $z_t(u)\geq\rho$, with our existing  monotonicity arguments ensuring that
these labels form a prefix that can only grow over time.
We set cumulative orders equal to the passing prefix length and place a new order corresponding to the increased length since the preceding period. The box below formalizes this rule, expressing the prefix length
as an integral of its membership indicator. In Appendix~\ref{subsec:divisible-analysis}, we verify feasibility
and show how integrating the individual stopping comparisons
yields the desired approximation guarantee.

\begin{tcolorbox}[
    breakable,
    colback=white,
    colframe=black!65,
    boxrule=0.6pt,
    arc=0pt,
    left=9pt,
    right=9pt,
    top=8pt,
    bottom=8pt,
    before skip=12pt,
    after skip=12pt
]
\textbf{The modified common-threshold policy
{\boldmath $\policydiv$}}

\medskip
Follow the common-threshold policy of
Section~\ref{subsec:construction}, sampling an independent threshold
$\rho$ for the entire planning horizon and initializing
$Q_{[0]}=0$. At the beginning of each ordering period $t\in[n]$,
replace its search over integer labels by the following rule:
\begin{enumerate}
    \item \textbf{Determine cumulative orders.}
    Set the cumulative order quantity equal to the length of
    the passing prefix, i.e., $Q_{[t]} = \int_0^M \ind\{z_t(u)\geq\rho\}\dd u$.
    Here, the accumulated clocks are evaluated based on the conditional charges at their
    original observed histories.

    \item \textbf{Place new order.}
    Implement the increase in cumulative supply by placing an order of size $Q_t
     = Q_{[t]}-Q_{[t-1]}$.
\end{enumerate}
\end{tcolorbox}

\subsection{Analysis}
\label{subsec:divisible-analysis}

\paragraph{Ordering feasibility.} Similarly to the integer setting, the pairwise comparisons of Lemma~\ref{lem:prefix} apply to real labels, meaning that $z_t(u)$ is nonincreasing in $u$ and nondecreasing in $t$. The passing sets are therefore growing prefixes, implying that $Q_{[t]} \geq Q_{[t-1]} \geq 0$, or put differently, every order quantity $Q_t = Q_{[t]} - Q_{[t-1]}$ is nonnegative. Once again, the accumulated clocks and their crossing indicators are jointly measurable in the available information. Integrating these indicators over labels makes $Q_{[t]}$ measurable with respect to
$\Hist_t\vee\sigma(\rho)$, so our policy is indeed nonanticipatory.

\paragraph{Continuous cost decomposition.} Moving forward, our objective is to bound the expected total cost of $\policydiv$ in relation to the fully adaptive benchmark $\opt_{\rm div}$. To this end, for any policy $P\in\PP_{\rm div}$, let us define the ordering time of every label $u\in(0,M]$ exactly as in Section~\ref{subsec:labels}, namely, $S_P(u) = \min( \{ t \in [n] : Q_{[t]}^P \geq u \} \cup \{n+1\} )$. For each fixed label, $S_P(u)$ is a stopping time with respect to the current history enlarged by the policy's independent seed. In addition, it is jointly measurable in the realization, seed, and label,
since $\{S_P(u)\leq t\}=\{Q_{[t]}^P\geq u\}$.

Mimicking earlier observations regarding our end-of-period-$t$ net inventory $I_t^P$, given by equations~\eqref{eq:holdunits}
and~\eqref{eq:backunits}, its continuous counterpart is now measured by interval lengths rather
than by counting labels, in the sense that
\begin{align*}
[I_t^P]^+ ~~&=~~ \int_0^M\ind\{D_{[t]}<u\} \cdot \ind\{S_P(u)\leq t-L\}\dd u, \\
[-I_t^P]^+  ~~&=~~\int_0^M\ind\{D_{[t]}\geq u\} \cdot \ind\{S_P(u)>t-L\}\dd u. 
\end{align*}
Since the fixed-label conditioning argument of Lemma~\ref{lem:unitcost}
applies unchanged, including for randomized policies, integration along with Tonelli's theorem yields an expected cost decomposition of the form
\begin{equation}\label{eq:unitcost-divisible}
\exsub{\omega,\xi_P}{F(P)} ~~=~~ K_0+\int_0^M\exsub{\omega,\xi_P}{ \alpha_{S_P(u)}(u)+ \sum_{t<S_P(u)} \beta_t(u)} \dd u,
\end{equation}
which is a continuous analogue of Lemma~\ref{lem:unitcost}, with the finite summation over $1, \ldots, M$ replaced by an integral over $[0,M]$. 

\paragraph{Applying the weighted stopping theorem.} To utilize this accounting scheme, it is worth pointing out that, fixing any comparison policy $P\in\PP_{\rm div}$ and real-valued label $u\in(0,M]$, the stopping prices  $\alpha_t(u)$ and waiting charges $\beta_t(u)$ remain adapted and bounded. Moreover, the conditioning argument of
Lemma~\ref{lem:supermart} applies without modification, showing that
$(\alpha_t(u))_{t\in[n+1]}$ is a nonnegative supermartingale.
Thus, the hypotheses behind Theorem~\ref{thm:stopping} hold for
each fixed real label, as in the integer model. As such, let $\tau_u(\rho) = \min( \{ t \in [n] : z_t(u) \geq \rho \} \cup \{n+1\} )$
be the stopping time prescribed by the common threshold $\rho$ for label $u$. By applying Theorem~\ref{thm:stopping} to this label, it follows that
\begin{equation}\label{eq:divisible-label-comparison}
 \exsub{\omega,\rho}{      \alpha_{\tau_u(\rho)}(u)    +\sum_{t<\tau_u(\rho)}\beta_t(u)}
 ~~\leq~~ \frac{e}{e-1} \cdot  \exsub{\omega,\xi_P}{      \alpha_{S_P(u)}(u)      +\sum_{t<S_P(u)}\beta_t(u)}.
\end{equation}

\paragraph{Approximation guarantee.}
We now integrate these comparisons over all continuous labels.
With respect to the common-threshold policy $\policydiv$, the crossing times $\tau_u(\rho)$ agree with our ordering times for almost every label
on each realization, and by combining~\eqref{eq:divisible-label-comparison}
with~\eqref{eq:unitcost-divisible}, we obtain
\begin{align*}
 \exsub{\omega,\rho}{F(\policydiv)}
 ~~&=~~
 K_0+
 \int_0^M
 \exsub{\omega,\rho}{
     \alpha_{\tau_u(\rho)}(u)
     +\sum_{t<\tau_u(\rho)}\beta_t(u)}\dd u
 \\
 &\leq~~
 K_0+\frac{e}{e-1}\cdot
 \int_0^M
 \exsub{\omega,\xi_P}{
     \alpha_{S_P(u)}(u)
     +\sum_{t<S_P(u)}\beta_t(u)}\dd u
 \\
 &=~~
 K_0+\frac{e}{e-1}\cdot
 \left(\exsub{\omega,\xi_P}{F(P)}-K_0\right)
 \\
 &\leq~~
 \frac{e}{e-1}\cdot
 \exsub{\omega,\xi_P}{F(P)}.
\end{align*}
Since this inequality holds for every comparison policy $P\in\PP_{\rm div}$,
taking the infimum over these policies yields $\exsubpar{\omega,\rho}{F(\policydiv)} \leq \frac{e}{e-1}\cdot\opt_{\rm div}$.

\end{document}